\documentclass{article}
\pdfoutput=1

\usepackage[utf8]{inputenc}
\usepackage[T1]{fontenc}

\usepackage{geometry}
\usepackage{color}

\usepackage{amsfonts}
\usepackage{bm}

\usepackage[tracking=true,kerning=true,final]{microtype}

\usepackage[sorting=none,url=false,isbn=false,doi=false,firstinits=true,maxbibnames=20,backend=bibtex]{biblatex}
\usepackage{caption}
\usepackage{subcaption}

\usepackage{graphicx}
\graphicspath{{./figs/}}

\usepackage{mathtools}
\mathtoolsset{showonlyrefs}

\usepackage[unicode=true,pdfusetitle, bookmarks=false, breaklinks=false,hidelinks,colorlinks=true]{hyperref}
\hypersetup{colorlinks=true, linkcolor=blue,  anchorcolor=blue, citecolor=blue, filecolor=blue, menucolor=blue, urlcolor=blue}
\hypersetup{pdfauthor={Mustafa A. Mohamad and Themistoklis P. Sapsis},pdftitle={Probabilistic description of extreme events in intermittently unstable systems excited by correlated stochastic processes},pdfsubject={Analytical characterization of heavy-tail statistics related to rare events in intermittently unstable dynamical systems.}}

\usepackage{lmodern}
\newcommand{\overbar}[1]{\mkern 2mu\overline{\mkern-2mu#1\mkern-2mu}\mkern 2mu}
\DeclarePairedDelimiter\abs{\lvert}{\rvert}

\DeclareMathOperator{\var}{\text{Var}}
\DeclareMathOperator{\prob}{\mathcal P}
\DeclareMathOperator{\real}{\text{Re}}

\title{Probabilistic description of extreme events in intermittently unstable systems excited by correlated stochastic processes}

\author{Mustafa A. Mohamad and Themistoklis P. Sapsis%
\thanks{Corresponding author: \href{mailto:sapsis@mit.edu}{sapsis@mit.edu},
Tel: (617) 324-7l508, Fax: (617) 253-8689%
}\\
Department of Mechanical Engineering,\\
Massachusetts Institute of Technology, \\
77 Massachusetts Ave., Cambridge, MA 02139}
\date{July 18, 2014}

\begin{document}

\maketitle
\begin{abstract}
In this work, we consider systems that are subjected to intermittent instabilities due to external stochastic excitation. These intermittent instabilities, though rare, have a large impact on the probabilistic response of the system and give rise to heavy-tailed probability distributions. By making appropriate assumptions on the form of these instabilities, which are valid for a broad range of systems, we formulate a method for the analytical approximation of the probability distribution function (pdf) of the system response (both the main probability mass and the heavy-tail structure). In particular, this method relies on conditioning the probability density of the response on the occurrence of an instability and the separate analysis of the two states of the system, the unstable and stable state. In the stable regime we employ steady state assumptions, which lead to the derivation of the conditional response pdf using standard methods for random dynamical systems. The unstable regime is inherently transient and in order to analyze this regime we characterize the statistics under the assumption of an exponential growth phase and a subsequent decay phase until the system is brought back to the stable attractor. The method we present allows us to capture the statistics associated with the dynamics that give rise to heavy-tails in the system response and the analytical approximations compare favorably with direct Monte Carlo simulations, which we illustrate for two prototype intermittent systems: an intermittently unstable mechanical oscillator excited by correlated multiplicative noise and a complex mode in a turbulent signal with fixed frequency, where multiplicative stochastic damping and additive noise model interactions between various modes.
\end{abstract}

\section{Introduction}

A wide range of dynamical systems describing physical and technological processes are characterized by intermittency, i.e. the property of having sporadic responses of extreme magnitude when the system is ``pushed'' away from its statistical equilibrium. This intermittent response is usually formulated through the interplay of stochastic excitation, which can trigger internal system instabilities, deterministic restoring forces (usually in terms of a potential) and dissipation terms. The result of intermittent instabilities can be observed by the heavy-tails in the statistics of the system response. It is often the case that, despite the high dimensionality of the stable attractor, where the system resides most of the time, an extreme response of short duration is due to an intermittent instability occurring over a single mode. This scenario does not exclude the case of having more than one intermittent mode, as long as the extreme responses of these modes are statistically independent. For this case, it may be possible to analytically approximate the probabilistic structure of these modes and understand the effect of the unstable dynamics on the heavy-tails of the response.

Instabilities of this kind are common in dynamical systems with uncertainty. One of the most popular examples are modes in turbulent fluid flows and nonlinear water waves subjected to nonlinear energy exchanges that occur in an intermittent fashion and result in intermittent responses \cite{Pedlosky_book1,salmonbook,Delsole,majda_info,majda1997,Dysthe08,xiao13}. Prototype systems that mimic these properties were introduced in \cite{majda_branicki_DCDS,branic_majda,Majda_filter}. In recent works, it has been shown that these properly designed single mode models can describe intermittent responses, even in very complex systems characterized by high dimensional attractors \cite{Majda_filter,cousins_sapsis,chen_majda_giannakis}. Another broad class of such systems includes mechanical configurations subjected to parametric excitations, such as parametric resonance of oscillators and ship rolling motion \cite{nayfeh_mook,Broer,Shadman,vakakis_kourdis,arnold_l}. Finally, intermittency can often be found in nonlinear systems that contain an invariant manifold that locally loses its transverse stability properties, e.g. in the motion of finite-size particles in fluids, but also in biological and mechanical systems with slow-fast dynamics \cite{babiano00,sapsishaller08,HallerSapsis09,poggiale}.

In all of these systems, the complexity of the unstable dynamics is often combined with stochasticity introduced by persistent instabilities that lead to chaotic dynamics as well as by the random characteristics of external excitations. The structure of the stochasticity introduced by these factors plays an important role on the underlying dynamics of the system response. In particular, for a typical case the stochastic excitation is colored noise, i.e. noise with finite correlation time length. Due to the possibility of large excursions by the colored stochastic excitation from its mean value into an ``unsafe-region'', where hidden instabilities are triggered, extreme events may be particularly severe. Therefore, for an accurate description of the probabilistic system dynamics, it is essential to develop analytical methods that will be able to capture accurately the effects of this correlation in the excitation processes for intermittent modes. However, analytical modeling in this case is particularly difficult since even for low-dimensional systems, standard methods that describe the pdf of the response (such as the Fokker-Planck equation) are not available.

For globally stable dynamical systems numerous techniques have been developed to analyze extreme responses (see e.g. \cite{naess_book,Soong_Grigoriou93,Naess1982}). There are various steps involved in this case that lead to elegant and useful results, but the starting point is usually the assumption of stationarity in the system response, which is not the case for intermittently unstable systems. Extreme value theory \cite{extreme_value09,Leadbetter,Thomas_extremes,Embrechts12,Galambos} is also a widely applicable method which focuses on thoroughly analyzing the extreme properties of stationary stochastic processes following various distributions. However, even in this case the analysis does not take into account any information about the system dynamics and is usually restricted to very specific forms of correlation functions \cite{extreme_value09,Leadbetter}.

Approaches that give more emphasis on the dynamics include averaging of the governing equations. However, the inherently transient character of intermittent instabilities makes it impossible for averaging techniques to capture their effect on the response statistics. Other modeling attempts include approximation of the correlated parametric stochastic excitation by white noise (see e.g. \cite{Kreuzer} for an application to parametric ship rolling motion). However, even though this idealization considerably simplifies the analysis and leads to analytical results, it underestimates the intensity of extreme events, since instabilities occur for an infinitesimal amount of time (the process spends an infinitesimal time in the unstable regime, which is insufficient time for the system to depart from the stable attractor). For this case, heavy-tails can be observed only as long as the parametric excitation is very intense, which for many systems is an unrealistic condition.

\emph{In this work, our goal is the development of a method that will allow for the analytical approximation of the probability distribution function of modes associated with intermittent instabilities and extreme responses due to parametric excitation by colored noise. This analytic approach will provide a direct link between dynamics and response statistics in the presence of intermittent instabilities.} We decompose the problem by conditioning the response of the system on the occurrence of unstable and stable dynamics. This idea enables the separate analysis of the system response in the two regimes and allows us to capture accurately the heavy-tails that arise due to intermittent instabilities. The full probability distribution of the system state is then reconstructed by combining the results from the two regimes. We illustrate this approach to two prototype systems that arise in problems of turbulent flows and nonlinear waves, as well as mechanics. Finally, we thoroughly examine the extent of validity of the derived approximations by direct comparison with Monte Carlo simulations.

\section{Problem setup and method}\label{sec:methodology}

Let $(\Theta,\mathcal{B},\mathcal{P})$ be a probability space, where $\Theta$ is the sample space with $\theta\in\Theta$ denoting an elementary event of the sample space, $\mathcal{B}$ the associated $\sigma$-algebra of the sample space, and $\mathcal{P}$ a probability measure. In this work we are interested in describing the statistical characteristics of modes subjected to intermittent instabilities. We consider a general dynamical system
\begin{equation}
 \bm{\dot{x}}= G (\bm{x},t), \quad \bm{x}\in\mathbb{R}^n.
\end{equation}

The presented analysis will rely on the following assumptions related to the form of the extreme events:
\begin{enumerate}
\item[A1] The instabilities are rare enough so that they can be considered statistically independent and have finite duration.
\item[A2] During an extreme event the influenced modes have decoupled dynamics. Moreover, for each one of these modes, during the growth phase, the instability is the governing mechanism, i.e. nonlinear terms are not considered to be important during these fast transitions, but only during the stable dynamics.
\item[A3] After each extreme event there is a relaxation phase that brings the system back to its stable stochastic attractor.
\end{enumerate}
Under these assumptions we may express each intermittent mode, denoted by $u(t;\theta)\in\mathbb{R}$, where $\theta\in\Theta$ (for notational simplicity, we drop explicit dependence on the probability space for random processes whenever clear), as a dynamical system of the form
\begin{equation}\label{eq:general_mode_interm}
\dot{u} + \alpha(t;\theta) u+\varepsilon\zeta(u,\bm{v})=\varepsilon\xi(t;\theta),
\end{equation}
where $\varepsilon>0$ is a small quantity and $\zeta(u,\bm{v})$ is a nonlinear function with zero linearization with respect to $u$, which may also depend on other system variables $\bm{v}\in\mathbb{R^{\mathrm{n-1}}}$. Since $\varepsilon$ is a small quantity, we can assume that the nonlinear term is important only in the stable regime. The stochastic processes $\alpha(t;\theta)$ and $\xi(t;\theta)$ are assumed stationary with known statistical characteristics. For $\alpha(t;\theta)$ we will make the additional assumption that its statistical mean is positive i.e. $\bar{\alpha}>0$ so that the above system has a stable attractor. The above equation can be thought as describing the modulation envelope for a complex mode with a narrow band spectrum, or one of the coordinates that describe the transverse dynamics to an invariant slow manifold.

\begin{figure}
\centering
\includegraphics[width=0.60\textwidth]{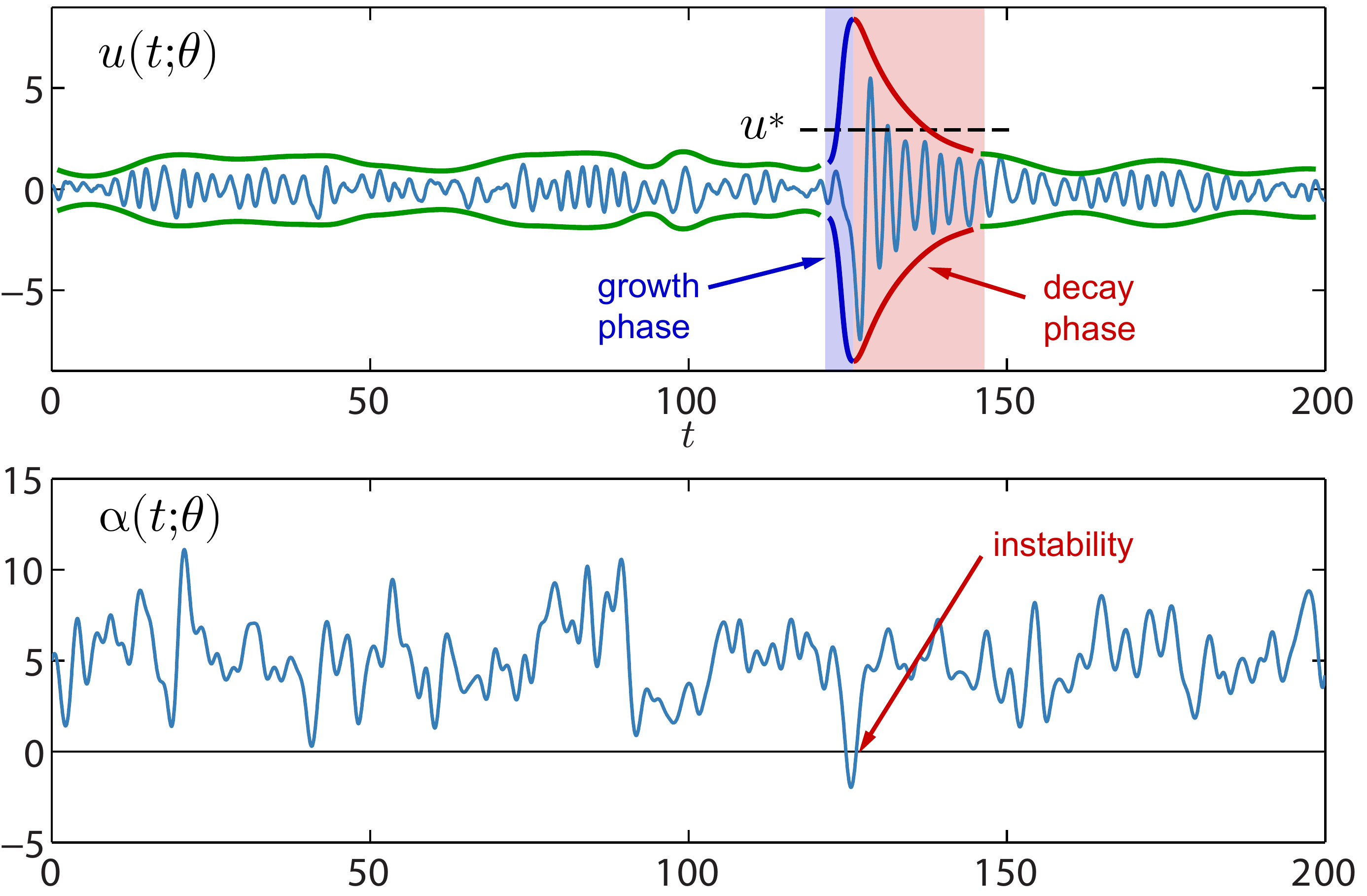}
\caption{An intermittently unstable system. Green lines denote the envelope
of the process during the stable regime. The blue lines correspond to the growth phase of the instability and the red lines to the relaxation phase of the instability.}
\label{fig:Sample_instability}
\end{figure}

The objective of this work is to derive analytical approximations for the probability distribution function of the system response, taking into account intermittent instabilities that arise due to the effect of the stochastic process\textit{ $\alpha(t)$}. In particular, these instabilities are triggered when $\alpha(t)<0$, and force the system to depart from the stable attractor. Therefore, the system has two regimes where the underlying dynamics behave differently: the stable regime where $\alpha(t)>0$ and the unstable regime that is triggered when $\alpha(t)<0$ (Fig. \ref{fig:Sample_instability}). Motivated by this behavior, we quantify the system's response by conditioning the probability distribution of the response on stable regimes and unstable events, and reconstruct the full distribution using Bayes' rule
\begin{equation}\label{eq:bayes}
\mathcal{P}[u]=\mathcal{P}[u\mid\text{stable regime}]\mathcal{P}[\text{stable regime}]+\mathcal{P}[u\mid\text{unstable regime}]\mathcal{P}[\text{unstable regime}],
\end{equation}
thereby separating the two regions of interest so that they can be individually studied. The method we employ relies on the derivation of the probability distribution for each of the terms in~\eqref{eq:bayes} and then the reconstruction of the full distribution of the system according to Bayes' rule. This allows us to capture the dramatically different statistics that govern the system response in the two regimes, since they are now decoupled. Essentially, we decouple the response pdf into two parts: a probability density function with rapidly decaying tails (typically Gaussian) and a heavy tail distribution with very low probability close to zero. In the following subsections, we provide a description of the method involved for the statistical determination of the terms involved in~\eqref{eq:bayes}.

\subsection{Stochastic description of the stable regime}\label{sec:methodologystableregim}

During the stable regime we have by definition $\alpha>0$ and therefore under this condition the considered mode is stable. Note that this condition is also true during the relaxation (decay) phase after an extreme event, when the system has entered a regime where $\alpha>0$. To this end, we cannot directly relate the duration of being on the stable attractor with the probability $\mathcal{P}[\alpha>0]$, but a correction should be made. We present this correction later in Section~\ref{sec:decayPhase}. Here we focus on characterizing the probability density function of the system under the assumption that it has relaxed to the stable attractor and moreover we have $\alpha>0$.

As a first-order estimate of the stable dynamics we approximate the original dynamical system for the intermittent mode by the stable system
\begin{equation}\label{eq:stable_general}
\dot{u}+ \bar{\alpha}|_{\alpha>0}u + \varepsilon \zeta(u,\bm{v})=\varepsilon\xi(t;\theta),
\end{equation}
where $\bar{\alpha}|_{\alpha>0}$ denotes the conditional average of the process $\alpha(t;\theta)$ given that this is positive. The determination of the statistical structure of the stable attractor for~\eqref{eq:stable_general} can be done with a variety of analytical and numerical methods, such as the Fokker-Planck equation if the process $\xi$ is white noise (see e.g.~\cite{Sobczyk91,Soong_Grigoriou93}) or the joint response-excitation equations otherwise~\cite{Sapsis08,venturi_sapsis}. Using one of these methods we can obtain the probability distribution function for the statistical steady state of the system $\mathcal{P}[u\mid\text{stable regime}]$.

\subsection{Stochastic description of the growth phase}

In contrast to the stable regime, the unstable regime is far more complicated due to its inherently transient nature. In addition, the unstable regime consists of two distinct phases: a growth phase where $\alpha<0$ and a decay phase where $\alpha>0$. We first consider the growth phase, where we rely on assumption~A2, according to which the dominant mechanism is the term related to the instability. Under this assumption, the first-order approximation during the growth phase is governed by the system
\begin{equation}\label{eq:growth_phase}
\dot{u}+\alpha(t;\theta)u = 0 \implies u(t;\theta)=u_{0}e^{\varLambda T},
\end{equation}
where $u_{0}$ is a random initial condition described by the probability measure in the stable regime, $T$ is the random duration of the upcrossing event $\alpha<0$, and $\varLambda$ is the growth exponent, which for each extreme event can be approximated by
\begin{equation}
\varLambda\simeq-\frac{1}{T}\int_{T}\alpha(t;\theta)\,dt\simeq-\alpha(t;\theta),
\end{equation}
due to the rapid nature of the growth phase. Therefore, during the growth phase we have (for notational clarity, we use the notation for the probability measure $\mathcal P$ to also represent the cumulative probability measure for suitable expressions)
\begin{equation}\label{eq:probab_unstable_basic}
\mathcal{P}[u>u^{*} \mid \alpha<0] = \mathcal{P}[u_{0}e^{\varLambda T}>u^{*}\mid\alpha<0] = \mathcal{P}[u_{0}e^{\varLambda T}>u^{*}\mid\alpha<0,u_{0}]\mathcal{P}[u_{0}].
\end{equation}
The right hand side of~\eqref{eq:probab_unstable_basic} is a derived distribution depending on the probabilistic structure of $\varLambda$ and $T$. The initial value $u_{0}$ is a random variable with statistical characteristics corresponding to the stable regime of the system, i.e. by the probability distribution function $\mathcal{P}[u\mid\text{stable regime}]$. Hence, to determine the required probability distribution we need only to know $\mathcal{P}[\alpha,T\mid\alpha<0],$ i.e. the joint probability distribution function for the value of $\alpha$ (given that this is negative) and the duration of the time interval over which $\alpha$ is negative. This distribution involves only the excitation process $\alpha$, and for the Gaussian case it can be approximated analytically (see Section~\ref{sec:gausproc}). Alternatively, one can compute this distribution using numerically generated random realizations that respect the statistical characteristics of the process.

\subsection{Stochastic description of the decay phase}\label{sec:decayPhase}

The decay phase is also an inherently transient stage. It occurs right after the growth phase of an instability, when $\alpha$ has an upcrossing of the zero level, and is therefore characterized by positive values of $\alpha$, with the effect of driving the system back to the stable attractor. To provide a statistical description for the relaxation phase, we first note the strong connection between the growth and decay phase. In particular, as shown in Fig.~\ref{fig:Sample_instability}, for each extreme event there is a one-to-one correspondence for the values of the intermittent variable $u$ between the growth phase and the decay phase. By focusing on an individual extreme event, we note that the probability of $u$ exceeding a certain threshold during the growth phase is equal with the probability of $u$ exceeding the same threshold during the decay phase. Thus over the total instability have
\begin{equation}\label{eq:cond_unstable}
\mathcal{P}[ u > u^{*} \mid \text{unstable regime}]= \mathcal{P}[u>u^{*}\mid\text{instability -- decay}] = \mathcal{P}[u>u^{*}\mid \text{instability -- growth}],
\end{equation}
where the conditional distribution for the growth phase has been determined in~(\ref{eq:probab_unstable_basic}).

\subsection{Probability of the stable and the unstable regime}

In the final step we determine the relative duration of the stable and unstable regimes. This ratio will define the probability of a stable event and an unstable event. The probability of having an instability is simply $P[\alpha<0]$, however, due to the decay phase the duration of an instability will be longer than the duration of the event $\alpha<0$. To determine the typical duration of the decay phase, we first note that during the growth phase we have
\begin{equation}\label{eq:variation_growth}
u_{p} = u_{0} e^{-\bar\alpha|_{\alpha<0} T_{\alpha<0}},
\end{equation}
where $T_{\alpha<0}$ is the duration for which $\alpha<0$, and $u_{p}$ is the peak value of $u$ during the instability. Similarly, for the decay phase we utilize system~\eqref{eq:stable_general} and obtain
\begin{equation}\label{eq:variation_decay}
u_{0}=u_{p}e^{-\bar{\alpha}|_{\alpha>0} T_\text{decay}}.
\end{equation}
Combining the last two equations~\eqref{eq:variation_growth} and~\eqref{eq:variation_decay} we have
\begin{equation}\label{eq:Probability_connection}
\frac{T_{\alpha<0}}{T_\text{decay}} = -\frac{\bar{\alpha}|_{\alpha>0}}{\bar{\alpha}|_{\alpha<0}}.
\end{equation}
This equation~\eqref{eq:Probability_connection} expresses the typical ratio between the growth and the decay phase. Thus, the total duration of an unstable event is given by the sum of the duration of these two phases
\begin{equation}\label{eq:instab_duration}
T_\text{inst}=\biggl( 1 - \frac{\bar{\alpha}|_{\alpha<0}}{\bar{\alpha}|_{\alpha>0}}\biggr) T_{\alpha<0}.
\end{equation}
Using this result, we can express the total probability of being in an unstable regime by
\begin{equation}
\mathcal{P}[\text{unstable regime}]=\biggl(1 - \frac{\bar{\alpha}|_{\alpha<0}}{\bar{\alpha}|_{\alpha>0}}\biggr)\mathcal{P}[\alpha< 0 ].
\end{equation}
Note that since we have assumed in~A1 that instabilities are sufficiently rare so that instabilities do not overlap and that instabilities are statistical independent, we will always have $\mathcal{P}[\text{unstable regime}]<1.$ Hence, the corresponding probability of being in the stable regime is
\begin{equation}
\mathcal{P}[\text{stable regime}]=1-\biggl(\mathrm{1-\frac{\bar{\alpha}|_{\alpha<0}}{\bar{\alpha}|_{\alpha>0}}}\biggr)\mathcal{P}[\alpha< 0 ].
\end{equation}

\section{Instabilities driven by Gaussian processes}\label{sec:gausproc}

In this section, we recall relevant statistical properties associated with a Gaussian stochastic parametric excitation. The zero level of the stochastic process $\alpha(t)$ defines the boundary of the two states for our system. The statistics that are relevant to our problem include the probability that the stochastic process is above and below the zero level, the average duration spent below and above the zero level, and the probability distribution of the length of time intervals spent below the zero level.

For convenience, let $\alpha(t;\theta)= m + k\gamma(t;\theta)$, where $\gamma(t)$ is also an ergodic and stationary Gaussian process, but with zero mean and unit variance, so that $\alpha(t)$ has mean $m$ and variance $k^{2}$. Thus the threshold of a rare event in terms of $\gamma(t)$ is given by the parameter $\eta \equiv -m/k$. We assume that second order properties, such as the power spectrum, of $\gamma(t)$ are known. In such a case, the correlation of the process is given by
\begin{equation}
R_{\gamma}(\tau)=\int_{-\infty}^{\infty}S_{\gamma}(\omega)e^{i\omega\tau}\,d\omega.
\end{equation}
In addition, we denote by $\phi(\,\cdot\,)$ the standard normal probability density function and by $\Phi(\,\cdot\,)$ the standard normal cumulative probability density function.

Since $\gamma(t)$ is a stationary Gaussian process, the probability that the stochastic process is in the two states $\mathcal{P}[\alpha<0]$ and $\mathcal{P}[\alpha>0]$ are, respectively, $\mathcal{P}[\gamma<\eta]=\Phi(\eta)$ and $\mathcal{P}[\gamma>\eta]=1-\Phi(\eta)$.

\subsection{Average time below and above the zero level}

Here we determine the average length of the intervals that $\alpha(t)$ spends above and below the zero level. For the case $\alpha(t)<0$, that is $\gamma(t)<\eta$, the expected number of upcrossings of this threshold per unit time is given by Rice's formula~\cite{MR0370729,MR2264709} (for a stochastic process with unit variance)
\begin{equation}\label{rice}
\overbar{N^{+}}(\eta)=\int_{0}^{\infty}u f_{\gamma\dot{\gamma}}(\eta,u)\,du  = \frac{1}{2\pi}\sqrt{-R_{\gamma}''(0)}\exp(-\eta^{2}/2),
\end{equation}
where $\overbar{N^{+}}(\eta)$ is the average number of upcrossings of level $\eta$ per unit time, which is equivalent to the average number of downcrossings $\overbar{{N}^{-}}(\eta)$. The expected number of crossings is finite if and only if $\gamma(t)$ has a finite second spectral moment~\cite{MR2264709}.

The average length of the interval that $\gamma(t)$ spends below the threshold $\eta$ can then be determined by noting that this probability is given by the product of the number of downcrossings of the threshold per unit time and the average length of the intervals for which $\gamma(t)$ is below the threshold $\eta$~\cite{MR0094269}
\begin{equation}
\bar{T}_{\alpha<0}=\frac{\mathcal{P}[\gamma<\eta]}{\overbar{{N}^{-}}(\eta)}.
\end{equation}
Hence, using the result~\eqref{rice} we have,
\begin{equation}\label{eq:meantimebelow}
\bar{T}_{\alpha<0}(\eta)=\frac{2\pi\exp(\eta^{2}/2)}{\sqrt{-R_{\gamma}''(0)}}\Phi(\eta).
\end{equation}

\subsection{Distribution of time below the zero level}

Here we recall a result regarding the distribution of time that the stochastic process $\gamma(t)$ spends below the threshold level $\eta$, i.e. the probability of the length of intervals for which $\gamma(t)<\eta$.

\begin{figure}[htb]
\centering
\includegraphics[width=\textwidth]{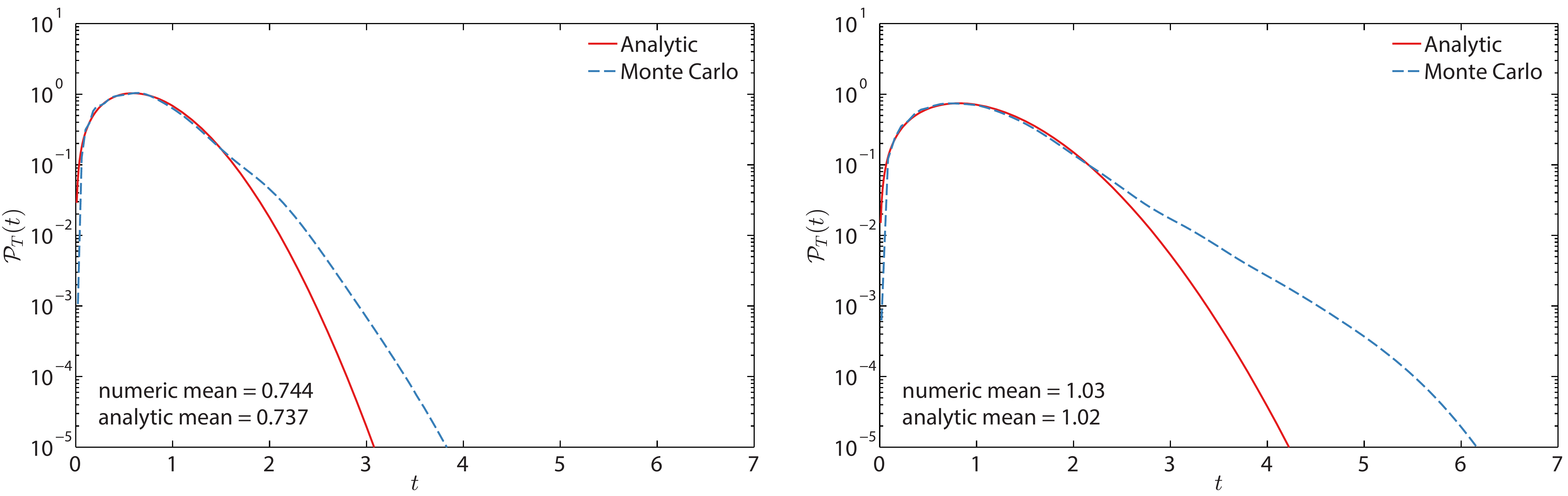}
\caption{Comparison of the analytic distribution $\mathcal{P}_{T}(t)$~\eqref{eq:rayleigh} with Monte Carlo simulation, generated using $1000$ ensembles of a Gaussian process with correlation $R(\tau)=\exp(-\tau^{2}/2)$ from $t_{0}=0$ to $t_{f}=1000$, and for two different sets of parameters: $m=5.0$, $k=1.6$ (left), $m=5.0$, $k=2.4$ (right).}
\label{timebelowdist}
\end{figure}

In general, it is not possible to derive an exact analytical expression for the distribution of time intervals given $\gamma(t)<\eta$, in other words the distribution of the length of time between a downcrossing and an upcrossing. However, the asymptotic expression in the limit $\eta\to-\infty$ is given by~\cite{MR0094269} (henceforth we denote $\bar{T}_{\alpha<0}$ by $\bar{T}$ for simplicity)
\begin{equation}\label{eq:rayleigh}
\mathcal{P}_{T}(t)=\frac{\pi t}{2\bar{T}^{2}}\exp\bigl(-{\pi t^{2}}/{4\bar{T}^{2}}\bigr),
\end{equation}
which is a Rayleigh distribution with scale parameter $\sqrt{2\bar{T}^{2}/\pi}$, and $\bar{T}$ is given by~\eqref{eq:meantimebelow}. This approximation is valid for small interval lengths $t$, since the derivation assumes that when a downcrossing occurs at time $t_{1}$ only a single upcrossing occurs at time $t_{2}=t_{1}+t$, neglecting the possibility of multiple crossings in between the two time instances. In Fig.~\ref{timebelowdist} the analytic distribution~\eqref{eq:rayleigh} is compared with numerical results, and as expected, we see that the analytic expression agrees well with the numerical results for small interval lengths, where the likelihood of multiple upcrossings in the interval is small. In addition, note that the only relevant parameter of the stochastic process $\gamma(t)$ that controls the scale of the distribution in equation~\eqref{eq:rayleigh} is the mean time spent below the threshold level, which depends only on the threshold $\eta$ and $R_{\gamma}''(0)$.

\section{Application to a parametrically excited mechanical oscillator}\label{sec:applicationoscillator}

Having formulated the general method, we  proceed to the first application, to that of a parametrically excited mechanical oscillator. More specifically, we consider the following single-degree-of-freedom oscillator, under parametric stochastic excitation and stochastic forcing,
\begin{equation}\label{eq:harmonic}
\ddot{x}(t;\theta)+c\dot{x}(t;\theta)+\kappa(t;\theta)x(t;\theta)=\sigma_{x}\xi(t;\theta),
\end{equation}
where $c$ and $\sigma_{x}$ are constants, $\kappa(t;\theta)$ is a stationary and ergodic Gaussian process of a given power spectrum $S_{\kappa}(\omega)$, and $\xi(t;\theta)$ denotes white noise. We may write equation~\eqref{eq:harmonic} in the state space form
\begin{align}
dx_{1}(t) & =x_{2}(t)\, dt\label{eq:harmonic2}\\
dx_{2}(t) & =-\bigl(cx_{2}(t)+\kappa(t)x_{1}(t)\bigr)\, dt+\sigma_{x}\, dW(t).
\end{align}

Our presentation will follow Section~\ref{sec:methodology}, although the analytical expressions for the pdf will be computed directly for position and velocity (instead of the envelope process $u$ as done in Section~\ref{sec:methodology}) for instructive purposes. At the end of this section we present comparisons with direct numerical simulations and thoroughly examine the limits of validity for the approximations we have made.

\subsection{Probability distribution in the stable regime}\label{sec:stable}

We first derive the probability distribution for the system under consideration given that the system is in the stable regime. Following Section~\ref{sec:methodologystableregim} we replace $\kappa(t)$ by the mean $\omega_{s}^{2}=\bar{\kappa}|_{\kappa>0}$. This approximation is valid since only small fluctuations of the process $\kappa(t)$ around the mean $\omega_{s}^{2}$ occur in this state, and these fluctuations have a minor impact on the system's probabilistic response.

Making this approximation~\eqref{eq:harmonic2} becomes
\begin{equation}\label{eq:stableapprox}
\ddot{x}(t)+c\dot{x}(t)+\omega_{s}^{2}x(t)=\sigma_{x}\xi(t)
\end{equation}
To calculate the mean $\omega_{s}^{2}$, note that its probability density function is given by $\mathcal{P}[\kappa=x\mid\kappa>0]=\phi((x-m)/k)/(k(1-\Phi(\eta))$, hence
\begin{equation}\label{eq:fastfreq}
\omega_{s}^{2}=\bar{\kappa}|_{\kappa>0}=m+k\frac{\phi(\eta)}{1-\Phi(\eta)}.
\end{equation}

The density in the stable regime can now be found by seeking a stationary solution of the Fokker-Planck equation of the form $\mathcal{P}(x_{1},x_{2})=\mathcal{P}(\mathcal{H})$, where $\mathcal{H}=\frac{1}{2}(\omega_{s}^{2}x_{1}^{2}+x_{2}^{2})$ is the total mechanical energy of the system. In this case the Fokker-Planck equation simplifies to~\cite{Soong_Grigoriou93}
\begin{equation}
c\mathcal{P}(\mathcal{H})+\frac{1}{2}\sigma_{x}^{2}\mathcal{P}'(\mathcal{H})=0 \implies \mathcal{P}(\mathcal{H})=q\exp\biggl(-\frac{2c}{\sigma_{x}^{2}}\mathcal{H}\biggr),
\end{equation}
where $q=c\omega_{s}/(\pi\sigma_{x}^{2})$ is a normalization constant. Taking the marginal with respect to $x_{2}$ gives the probability density for the system's position
\begin{equation}\label{eq:stablepos}
\mathcal{P}[x_{1}=x\mid\kappa>0]=\sqrt{\frac{c\omega_{s}^{2}}{\pi\sigma_{x}^{2}}}\exp\biggl(-\frac{c\omega_{s}^{2}}{\sigma_{x}^{2}}x^{2}\biggr),
\end{equation}
which is Gaussian with variance $\sigma^{2}=\sigma_{x}^{2}/(2c\omega_{s}^{2})$. Similarly, taking the marginal with respect to $x_{1}$ gives the probability density for the system's velocity
\begin{equation}\label{eq:stablevel}
\mathcal{P}[x_{2}=x\mid\kappa>0]=\sqrt{\frac{c}{\pi\sigma_{x}^{2}}}\exp\biggl(-\frac{c}{\sigma_{x}^{2}}x^{2}\biggr),
\end{equation}
where the variance is $\sigma^{2}=\sigma_{x}^{2}/(2c)$.

\subsection{Probability distribution in the unstable regime}\label{sec:unstable}

Here we derive the probability distribution for the system's position $x_{1}$ during the growth stage of an instability. In the unstable regime, the system initially undergoes exponential growth due to the stochastic process $\kappa(t)$ crossing below the zero level, which is the mechanism that triggers the instability. After a finite duration of exponential growth, which stops when $\kappa(t)$ crosses above the zero level, and due to the large velocity gradients that result, friction damps the response back to the stable attractor. Following assumption~A2, we consider the parametric excitation as the primary mechanism driving the instability during the growth phase, and therefore ignore the effect of friction during this stage of the instability.

\emph{Treating the system response as a narrow band process}, we describe the instability in terms of the envelope $u$ for the position variable by averaging over the fast frequency $\omega_{s}$. During the growth phase we have $u\simeq u_{0}(\theta)\exp(\Lambda(\theta)T_{\kappa<0}(\theta))$, where $u_{0}(\theta)$ is the value of the position's envelope before an instability (to be determined in Section~\ref{sec:envelope}). Note that this growth rate will be the same for both the position and the velocity variables.

\begin{figure}[htb]
\centering
\includegraphics[width=0.60\textwidth]{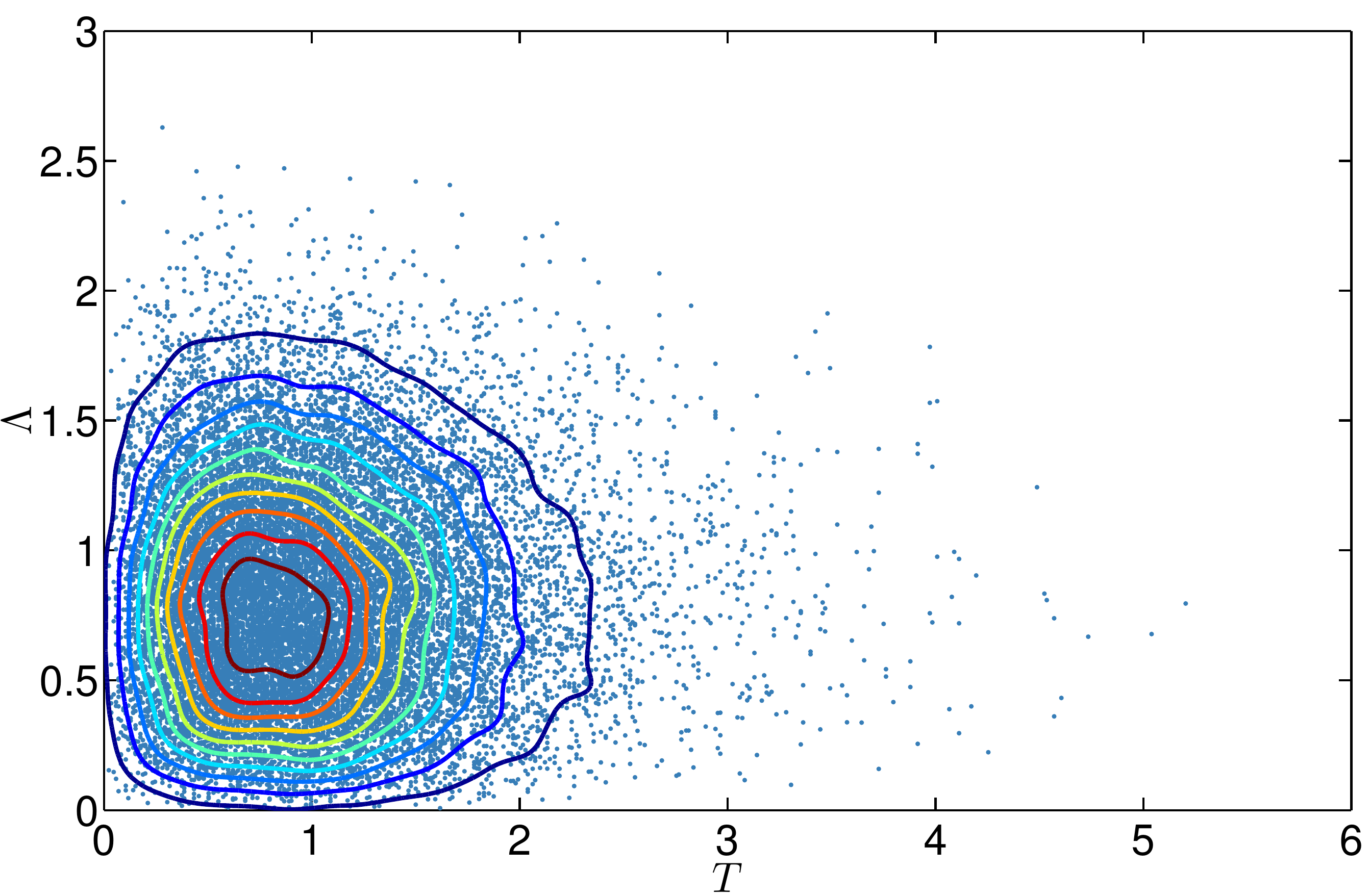}
\caption{Scatter plot and contours of the pdf of $T$ and $\Lambda$, for $\eta=-2.08$ ($m=5.00$, $k=2.40$). Samples for $\Lambda$ are drawn by inverse-sampling the cumulative distribution function and samples of $T$ are drawn from realizations of a Gaussian process using $1000$ ensembles of time length $t=1000$.}
\label{fig:corr}
\end{figure}

Substituting this representation into \eqref{eq:harmonic}, we find $\Lambda^{2}+\kappa\simeq0$, so that the positive eigenvalue is given by $\Lambda\simeq\sqrt{-\kappa}$. To proceed, we set $U=u_{0}\exp(\Lambda T)$ (for clarity denote $T_{\kappa<0}$ by $T$) and $Y=\Lambda$, then from a change of variables
\begin{equation}
\mathcal{P}_{UY}(u,y)=\mathcal{P}_{\Lambda T}(\lambda,t)|\det[\partial(\lambda,t)/\partial(u,y)]|.
\end{equation}
Next, we assume that $T$ and $\Lambda$ are independent which is reasonable as it can be seen from Fig.~\ref{fig:corr}. Then, we have
\begin{equation}
\mathcal{P}_{UY}(u,y)=\frac{\mathcal{P}_{\Lambda}(\lambda)\mathcal{P}_{T}(t)}{u_{0}\lambda\exp(\lambda t)}=\frac{1}{uy}\mathcal{P}_{\Lambda}(y)\mathcal{P}_{T}\biggl(\frac{\log(u/u_{0})}{y}\biggr),\quad u>u_{0},\; y>0.
\end{equation}
Taking the marginal density gives
\begin{equation}\label{eq:density1}
\mathcal{P}_{U}(u)=\frac{1}{u}\int_{0}^{\infty}\frac{1}{y}\mathcal{P}_{\Lambda}(y)\mathcal{P}_{T}\biggl(\frac{\log(u/u_{0})}{y}\biggr)\, dy.
\end{equation}
Now, since $\mathcal{P}[\Lambda]=\mathcal{P}[{\sqrt{-\kappa}|\kappa<0}]$
from another change of variables we obtain
\begin{equation}\label{eq:alphaden}
\mathcal{P}_{\Lambda}(\lambda)=\frac{2\lambda}{\Phi(\eta)}\mathcal{P}_{\kappa}(-\lambda^{2})=\frac{2\lambda}{k\Phi(\eta)}\phi\biggl(-\frac{\lambda^{2}+m}{k}\biggr),\quad\lambda>0,
\end{equation}
since $\mathcal{P}_{\kappa}$ is normaly distributed. Combining this result and the fact that $\mathcal{P}_{T}$ is given by the Rayleigh distribution ~\eqref{eq:rayleigh}, with equation ~\eqref{eq:density1} gives the final result for the probability density function
\begin{equation}\label{eq:fullden}
\mathcal{P}[u\mid\kappa<0,u_{0}]=\frac{\pi\log(u/u_{0})}{k\bar{T}^{2}\Phi(\eta)u}\int_{0}^{\infty}\frac{1}{y}\phi\biggl(-\frac{y^{2}+m}{k}\biggr)\exp{\bigg(-\frac{\pi}{4\bar{T}^{2}y^{2}}\log(u/u_{0})^{2}\bigg)}\,dy,\quad u > u_0.
\end{equation}
Moreover, utilizing the expression for the duration of a typical extreme event (equation~\eqref{eq:instab_duration}) we obtain for this particular system
\begin{equation}\label{eq:insta_dur}
T_{\text{inst}}=\biggl(1+\frac{2\bar{\Lambda}_{\kappa<0}}{c}\biggr)T_{\kappa<0},
\end{equation}
where for the relaxation phase we have the rate of decay for the envelope being  $-c/2.$

To formulate our result in terms of the position variable, we refer to the narrow band approximation made at the begining of this analysis. This will give approximately $x_{1}=u \cos\varphi$ where $\varphi$ is a uniform random variable distributed between 0 and $2\pi$. The probability density function for $z=\cos\varphi$ is given by $\mathcal{P}[z]= 1/(\pi\sqrt{1-z^{2}}), z\in [-1,1].$ To avoid double integrals for the computation of the pdf for the position, we approximate the pdf for $z$ by $\mathcal{P}[z]=\frac{1}{2}(\delta(z+1)+\delta(z-1)).$ This will give the following approximation for the conditionally unstable pdf
\begin{equation}\label{eq:eq_positio_unstable_oscillator}
\mathcal{P}[x_{1}=x\mid\kappa<0,u_{0}]=\frac{1}{2}\mathcal{P}[u= \abs{x} \mid \kappa<0,u_{0}].
\end{equation}

We note that as presented in Section~\ref{sec:methodology}, the pdf for the envelope $u$ may also be used for the decay phase. However, during the decay phase the oscillatory character (with frequency $\omega_{s}$) of the response has to also be taken into account. Nevertheless, to keep the expressions simple we will use equation~\eqref{eq:eq_positio_unstable_oscillator} to describe the pdf for position in the decay phase as well. We will show in the following sections that this approximation compares favorably with numerical simulations.

\subsubsection{Integral approximation for the conditionally unstable pdf}

\label{sec:asymp}

\begin{figure}[htb]
\centering \includegraphics[width=0.6\textwidth]{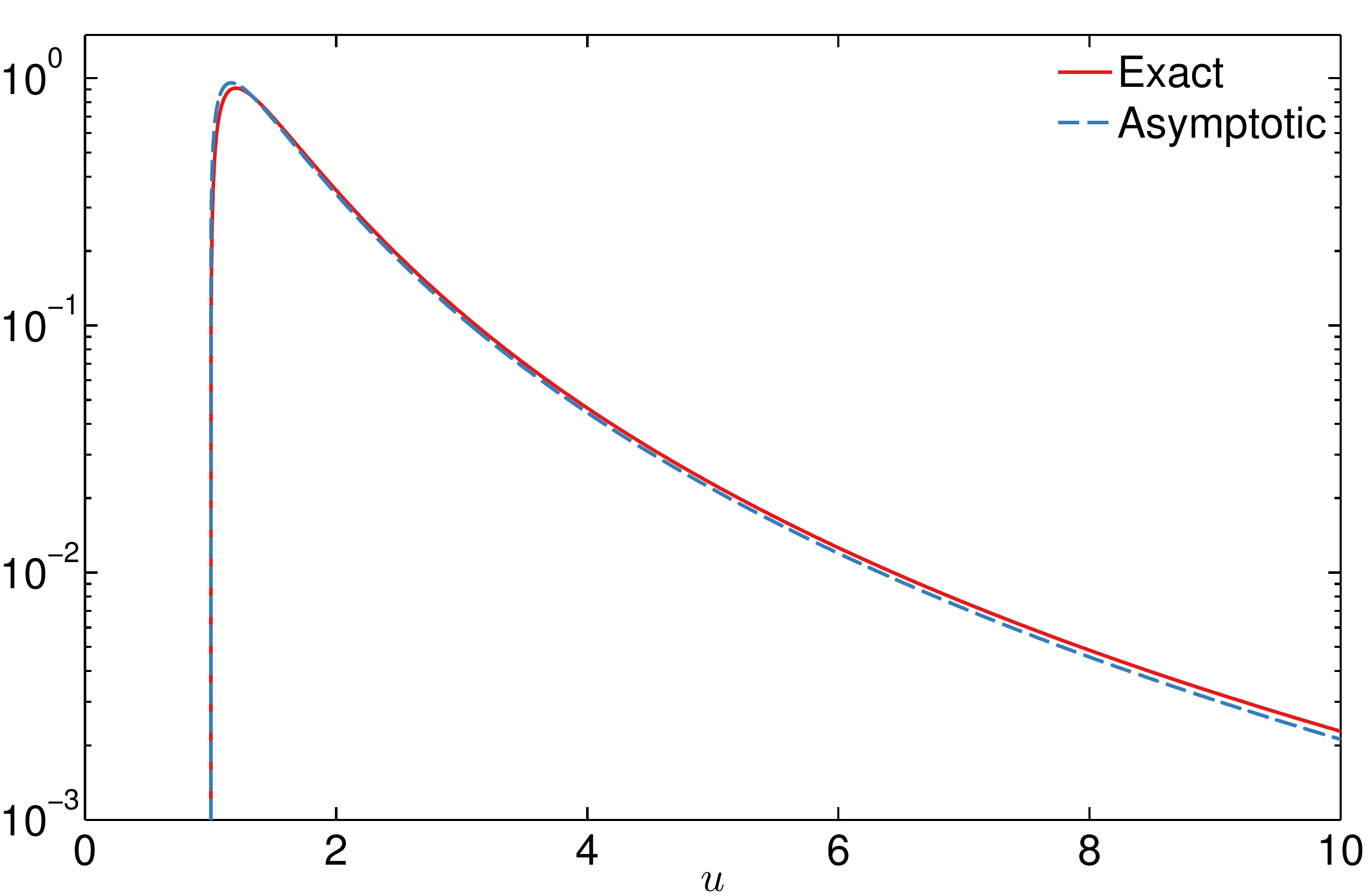}
\caption{Exact integral in equation~\eqref{eq:fullden} compared to the asymptotic expansion given in equation~\eqref{eq:asympunstab}, for a fixed $u_{0}$ and $m=5.00$, $k=2.00$.}
\label{fig:asympcompar}
\end{figure}

We derive an asymptotic expansion, as $u\to\infty$, for the integral~\eqref{eq:fullden} that describes the distribution for
the position's envelope during the growth phase of an instability,
treating $u_{0}$ as a parameter. For convenience in \eqref{eq:fullden} let
$\mathcal{P}[u\mid\kappa<0,u_{0}]=\int_{0}^{\infty}\xi(u,y;u_{0})\,dy$
that is
\begin{align}
\xi(u,y;u_{0}) & =\frac{\pi\log(u/u_{0})}{\sqrt{2\pi}k\bar{T}^{2}\Phi(\eta)uy}\exp\biggl(-\frac{(y^{2}+m)^{2}}{2k^{2}}-\frac{\pi}{4\bar{T}^{2}y^{2}}\log(u/u_{0})^{2}\biggr),\\
 & =\frac{\pi\log(u/u_{0})}{\sqrt{2\pi}k\bar{T}^{2}\Phi(\eta)uy}\exp(\theta(u,y;u_{0})),
\end{align}
where $\theta(u,y;u_{0})$ denotes the terms being exponentiated (we suppress the explicit dependence on the parameter $u_{0}$ for convenience), in other words
\begin{equation}
\mathcal{P}[u\mid\kappa<0,u_{0}]=\int_{0}^{\infty}\xi(u,y;u_{0})\, dy=\frac{\pi\log(u/u_{0})}{\sqrt{2\pi}k\bar{T}^{2}\Phi(\eta)u}\int_{0}^{\infty}\frac{1}{y}\exp(\theta(u,y))\, dy.
\end{equation}
This integral can be approximated following Laplace's method, where in this case we have a moving maxima~\cite{MR1721985}. For fixed $u$ the exponential is distributed around its peak value $y^{*}=\max_{y}\exp(\theta(u,y;u_{0}))$, so that we may make the following approximations to leading-order
\begin{align}
\mathcal{P}[u\mid\kappa<0,u_{0}] & \simeq\frac{\pi\log(u/u_{0})}{\sqrt{2\pi}k\Phi(\eta)u}\int_{y^{*}-\epsilon}^{y^{*}+\epsilon}\frac{1}{y^{*}}\exp\biggl(\theta(u,y^{*})+\frac{(y-y^{*})^{2}}{2}\partial_{yy}\theta(u,y^{*})+\dotsb\biggr)\, dy\\
 & \simeq\xi(u,y^{*};u_{0})\int_{-\infty}^{\infty}\exp\biggl(\frac{(y-y^{*})^{2}}{2}\partial_{yy}\theta(u,y^{*})+\dotsb\biggr)\, dy\\
 & =\frac{1}{\tilde{c}}\xi(u,y^{*};u_{0}),
\end{align}
where $\tilde{c}$ is a normalization constant that is computed numerically. In order to arrive at a closed analytic expression for $y^{*}=\max_{y}\exp(\theta(u,y))$, we find that the following expressions are good approximations for $y^{*}$ (see Appendix I for details)
\begin{align}
y_{1}^{*}(u;u_{0}) & =\biggl(\frac{\pi k^{2}}{4m\bar{T}^{2}}\biggr)^{1/4}\log(u/u_{0})^{1/2},\quad m^2\gg k^{2}\label{eq:maxy}\\
y_{2}^{*}(u;u_{0}) & =\biggl(\frac{\pi k^{2}}{4\bar{T}^{2}}\biggr)^{1/6}\log(u/u_{0})^{1/3},\quad m^2\ll k^{2}.
\end{align}
Hence, we have the following expressions that approximate the original integral
\begin{align}
\mathcal{P}[u\mid\kappa<0,u_{0}] & \simeq\xi(u,y_{1}^{*};u_{0}),\quad m^2\gg k^{2},\label{eq:asympunstab}\\
\mathcal{P}[u\mid\kappa<0,u_{0}] & \simeq\xi(u,y_{2}^{*};u_{0}),\quad m^2\ll k^{2},
\end{align}
where each expression needs to be suitably normalized. Equation~\eqref{eq:asympunstab}, in particular, is the most relevant in our case because the parameter $k$ that controls the frequency of instabilities will be smaller than the mean $m$ of the Gaussian process $\kappa(t)$, since our analysis is under the assumption that instabilities are sufficiently rare; in this case, note that the tail of the distribution decays like $\exp(\log(u/u_{0})^{1/2})$. In addition, the expansion in equation~\eqref{eq:asympunstab} matches the original integral extremely well, even for small values of $u$ (see Fig.~\ref{fig:asympcompar}).

\subsection{Probability density function for the envelope in the stable regime}\label{sec:envelope}

In the derivation of the probability density in the unstable regime we conditioned the result on the initial value of the position's envelope right before an instability occurs. The stable regime envelope is defined as the locus of local maxima for the stationary Gaussian process that governs the system before an instability. For a stationary Gaussian process of zero mean the joint probability density of the envelope and the envelope velocity is given by~\cite{Soong_Grigoriou93}
\begin{equation}\label{eq:maxgaus}
\mathcal{P}[u,\dot{u}]=\frac{u}{q\lambda_{0}\sqrt{2\pi\lambda_{2}}}\exp\biggl\{-\frac{1}{2}\biggl(\frac{u^{2}}{\lambda_{0}}+\frac{\dot{u}^{2}}{q^{2}\lambda_{2}}\biggr)\biggr\},
\end{equation}
where $\lambda_{n}$ is the $n$-th order spectral moment for the one-sided power spectral density of $x_{1}$ in the stable regime and $q^{2}=1-\lambda_{1}^{2}/\lambda_{0}\lambda_{2}$ describes the extent to which the process is narrow banded. In our case, we are interested in the pdf of the envelope for the position
\begin{equation}
\mathcal{P}[u]=\frac{u}{\lambda_{0}}\exp(-u^{2}/2\lambda_{0}),\label{eq:refenvray}
\end{equation}
which is a Rayleigh distribution with scale parameter $\sqrt{\lambda_{0}}$. The zeroth-order spectral moment is just the variance of the Gaussian pdf in~\eqref{eq:stablepos}, i.e. $\lambda_{0}=\sigma_{x}^{2}/(2c\omega_{s}^{2})$, and for convenience we write~\eqref{eq:refenvray} as
\begin{equation} \vartheta(u_{0})=\frac{2c\omega_{s}^{2}u_{0}}{\sigma_{x}^{2}}\exp(-c\omega_{s}^{2}u_{0}^{2}/\sigma_{x}^{2}).
\end{equation}
Therefore, we finally have the following result for the density in the unstable regime,
\begin{equation}
\mathcal{P}[x_{1}=x\mid\kappa<0]=\frac{1}{2}\mathcal{P}[u=|x|\mid\kappa<0]=\int_{0}^{\abs{x}}\frac{1}{2\tilde{c}}\xi(\abs{x},y_{1}^{*}(\abs{x};u_0);u_{0})\vartheta(u_{0})\, du_{0}.\label{eq:asympunstabdenpos}
\end{equation}

\subsection{Probability density function for the velocity}

\label{sec:velInter}

The probability density for the system's velocity $x_{2}$ in the stable regime can be found by marginalizing the solution obtained from the Fokker-Planck equation~\eqref{eq:stablevel}. The same result may be obtained by relying on the narrow band property of the stochastic response in the stable regime. In particular, we may represent the stochastic process for position $x_{1}$ in terms of its spectral power density ~\cite{Naess1982}:
\begin{equation}
x_{1}=\int_{0}^{\infty}\cos(\omega t+\varphi(\omega))\sqrt{2S_{x}(\omega)d\omega},
\end{equation}
where the last integral is defined in the mean square sense. Differentiating
the above we obtain
\begin{align}
x_{2} &=  \int_{0}^{\infty}\omega \cos\Bigl(\omega t+\varphi(\omega)+\frac{\pi}{2}\Bigr)\sqrt{2S_{x}(\omega)d\omega}\\
 & \simeq \omega_{s}\int_{0}^{\infty}\cos\Bigl(\omega t+\varphi(\omega)+\frac{\pi}{2}\Bigr)\sqrt{2S_{x}(\omega)d\omega},
\end{align}
using the narrow band property of the spectral density around $\omega_{s}$. Note that the integral in the last equation has the same distribution as $x_{1}$. Thus, it is easy to prove that the probability density function of $x_{2}$ in the \emph{stable regime} is given by the pdf of $\omega_{s}x_{1}.$

\begin{figure}
\centering
\includegraphics[width=0.6\textwidth]{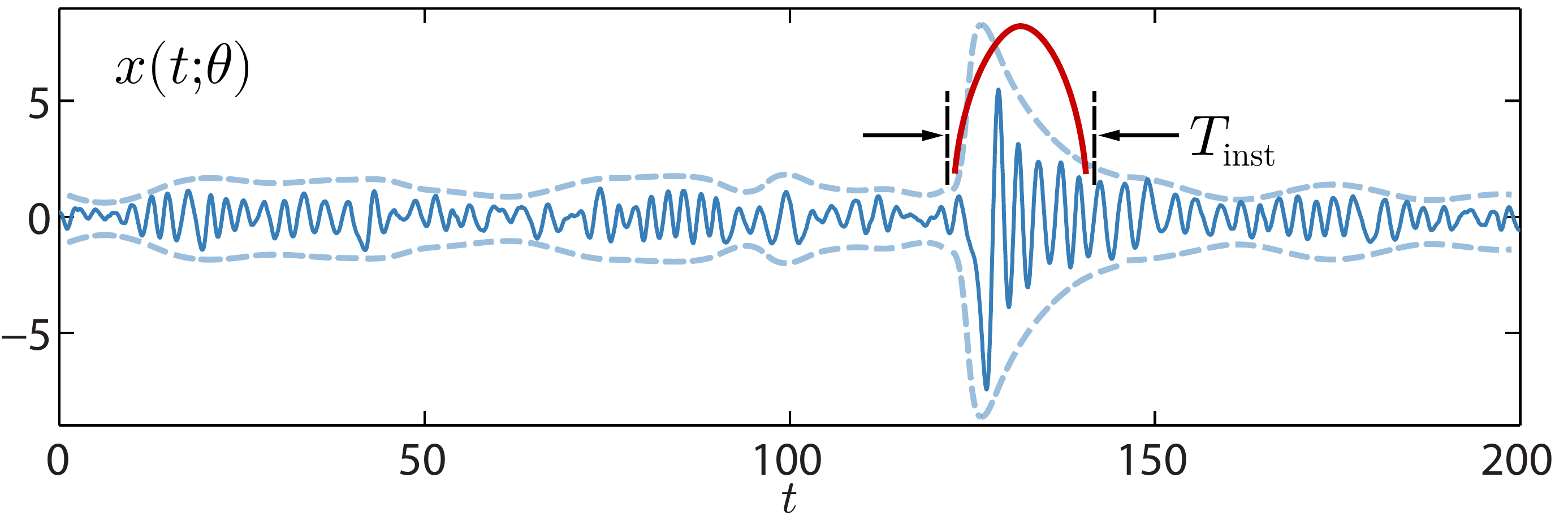}
\caption{For the computation of the velocity pdf, we approximate the envelope of the process during the extreme event with a harmonic function of consistent period.}
\label{Fig_vel_env}
\end{figure}

In the unstable regime stationarity breaks down, and the envelope
of the stochastic processes describing position changes with time.
This has an important influence on the probabilistic structure
for the velocity, which cannot be approximated as it was done in the
stable regime. The reason is that we have an additional time scale
associated with the variation of the envelope of the narrow-band oscillation
(see Fig.~\ref{Fig_vel_env}). For the purpose of computing the
pdf for velocity in the unstable regime, we approximate the envelope
for position during the instability by a sine function with a
half-period equal to the average duration of the extreme events  (equation~\eqref{eq:insta_dur}) $T_{\text{inst}}=(1+{2\bar{\Lambda}_{\kappa<0}}/{c})T_{\alpha<0}.$ This gives the following approximate expression for the unstable regime
\begin{equation}
x_{1}=\biggl(1+\frac{u_{p}}{u_{0}}\biggr)\sin(\omega_{u}t)\int_{0}^{\infty}\cos(\omega t+\varphi(\omega))\sqrt{2S_{x}(\omega)d\omega},
\end{equation}
where ${u_{p}}/{u_{0}}$ is the ratio of the extreme event peak value for the envelope over the initial value of the envelope (before the extreme event), and $\omega_{u}=\pi/T_{\text{inst}}$. Assuming that ${u_{p}}/{u_{0}}\gg1$ we  have that the pdf for velocity during the \emph{unstable regime} is also given in terms of $x_{1}$ but with the different scaling factor
\begin{equation}
x_{2}\sim\frac{1}{2}\biggl(\omega_{s}+\frac{\pi}{(1+ 2\bar{\Lambda}_{\kappa<0}/c )\bar{T}_{\alpha<0}}\biggr)x_{1}.
\end{equation}
The last relation gives a direct expression for the pdf of velocity in the unstable regime. Using this scaling factor   $\omega_{\text{inst}} =(\omega_s +\omega_u)/2 $ for the conditionally unstable pdf and the correct scaling for the stable regime $\omega_s$, we can derive an approximation for the pdf of velocity (given in the Section~\ref{sec:analytic}).

\subsection{Summary of results for the parametrically excited oscillator}

\label{sec:analytic}

\subsubsection*{Probability distribution function for the position}

Combining the results of equations~\eqref{eq:stablepos}, \eqref{eq:asympunstabdenpos}, and \eqref{eq:insta_dur} into the Bayes' decomposition in~\eqref{eq:bayes} gives the following heavy-tailed, symmetric probability density function for position
\begin{multline}
\mathcal{P}[x_{1}=x]=\bigl(1-(1+{2\bar{\Lambda}_{\kappa<0}}/c)\Phi(\eta)\bigr)\sqrt{\frac{c\omega_{s}^{2}}{\pi\sigma_{x}^{2}}}\exp\biggl(-\frac{c\omega_{s}^{2}}{\sigma_{x}^{2}}x^{2}\biggr)+\\
(1+{2\bar{\Lambda}_{\kappa<0}}/c)\Phi(\eta)\int_{0}^{\abs{x}}\frac{1}{2 \tilde{c}}\xi(\abs{x},y_{1}^{*}(\abs{x};x_{0});x_{0})\vartheta(x_{0})\, dx_{0}, \quad  x \in \mathbb R.\label{eq:pdfpos}
\end{multline}
Writing each term explicitly we  have
\begin{multline}
\mathcal{P}[ x_{1} = x]=\bigl(1-(1+{2\bar{\Lambda}_{\kappa<0}}/c)\Phi(\eta)\bigr)\sqrt{\frac{c\omega_{s}^{2}}{\pi\sigma_{x}^{2}}}\exp\Biggl(-\frac{c\omega_{s}^{2}}{\sigma_{x}^{2}}x^{2}\biggr)+(1+{2\bar{\Lambda}_{\kappa<0}}/c)\Phi(\eta)\\
\frac{\sqrt{2\pi}cw_{s}^{2}}{2 \tilde{c}\sigma_{x}^{2}k\bar{T}^{2}\Phi(\eta)}\int_{0}^{\abs{x}}\frac{\log(\abs{x}/x_{0})}{(\abs{x}/x_{0})y_{1}^{*}(\abs{x};x_{0})}\exp\biggl(-\frac{({y_{1}^{*}(\abs{x};x_{0})}^{2}+m)^{2}}{2k^{2}}-\frac{\pi\log(\abs{x}/x_{0})^{2}}{4\bar{T}^{2}{y_{1}^{*}}(\abs{x};x_{0})^{2}}-\frac{c\omega_{s}^{2}}{\sigma_{x}^{2}}x_{0}^{2}\biggr)\, dx_{0}.\label{eq:pdfposExplicit}
\end{multline}

\subsubsection*{Probability distribution function for the velocity}

To obtain the pdf for velocity we scale the probability distribution function for position in~\eqref{eq:pdfpos}. As noted in Section~\ref{sec:velInter}, the correct scaling for the first term in equation~\eqref{eq:bayes}, which represents the stable regime, is given by $\omega_{s}$, and for the second term, for the unstable regime, the scaling is given by $\omega_{\text{inst}}$. Applying the random variable transformation $\mathcal{P}[x_{2}=x]=\mathcal{P}[x_{1}=x/\omega]/\omega$ with the appropriate scaling for each of these two terms yields
\begin{multline}
\mathcal{P}[ x_{2} = x]=\bigl(1-(1+{2\bar{\Lambda}_{\kappa<0}}/c)\Phi(\eta)\bigr)\sqrt{\frac{c}{\pi\sigma_{x}^{2}}}\exp\biggl(-\frac{c}{\sigma_{x}^{2}}x^{2}\biggr)+\\
(1+{2\bar{\Lambda}_{\kappa<0}}/c)\Phi(\eta)\int_{0}^{\abs{x}/\omega_{\text{inst}}}\frac{1}{2\omega_{\text{inst}} \tilde{c}}\xi(\abs{x}/\omega_{\text{inst}},y_{1}^{*}(\abs{x}/\omega_{\text{inst}};x_{0}');x_{0}')\vartheta(x_{0}')\, dx_{0}', \quad  x \in \mathbb R.\label{eq:pdfvel1}
\end{multline}
For the integral in the second term in equation~\eqref{eq:pdfvel1}, making the substitution $x_{0}=\omega_{\text{inst}}x_{0}'$ we have
\begin{multline}
\mathcal{P}[x_{2}={x}\mid\kappa<0]=\frac{\sqrt{2\pi}c}{2 \tilde{c}\sigma_{x}^{2}k\bar{T}^{2}\Phi(\eta)}\frac{\omega_{s}^{2}}{\omega_{\text{inst}}^{2}}\int_{0}^{\abs{x}}\frac{\log(\abs{x}/x_{0})}{(\abs{x}/x_{0})y_{1}^{*}(\abs{x};x_{0})}\\
\exp\biggl(-\frac{({y_{1}^{*}(\abs{x};x_{0})}^{2}+m)^{2}}{2k^{2}}-\frac{\pi\log(\abs{x}/x_{0})^{2}}{4\bar{T}^{2}{y_{1}^{*}}(\abs{x};x_{0})^{2}}-\frac{c}{\sigma_{x}^{2}}\frac{\omega_{s}^{2}}{\omega_{\text{inst}}^{2}}x_{0}^{2}\biggr)\, dx_{0}.\label{eq:expsecondVel}
\end{multline}
Therefore rewriting equation~\eqref{eq:pdfvel1}, gives the following pdf for velocity
\begin{multline}
\mathcal{P}[ x_{2} = {x}]=\bigl(1-(1+{2\bar{\Lambda}_{\kappa<0}}/c)\Phi(\eta)\bigr)\sqrt{\frac{c}{\pi\sigma_{x}^{2}}}\exp\biggl(-\frac{c}{\sigma_{x}^{2}}x^{2}\biggr)+\\
(1+{2\bar{\Lambda}_{\kappa<0}}/c)\Phi(\eta)\int_{0}^{\abs{x}}\frac{1}{2 \omega_{\text{inst}}\tilde{c}}\xi(\abs{x},y_{1}^{*}(\abs{x};x_{0});x_{0})\vartheta(x_{0}/\omega_{\text{inst}})\, dx_{0},\label{eq:pdfvel}
\end{multline}
where the integral in the second term is explicitly given   in equation~\eqref{eq:expsecondVel}.

\subsection{Comparison with direct Monte Carlo simulations and range of validity}

Here we discuss the analytic results for the system response pdf given in Section~\ref{sec:analytic}, present comparisons with Monte Carlo simulations, and describe the effect of varying system parameters on the analytical approximations. In addition, we examine the region of validity of our results with respect to these parameters.

\begin{figure}[htb]
\centering
\includegraphics[width=0.60\textwidth]{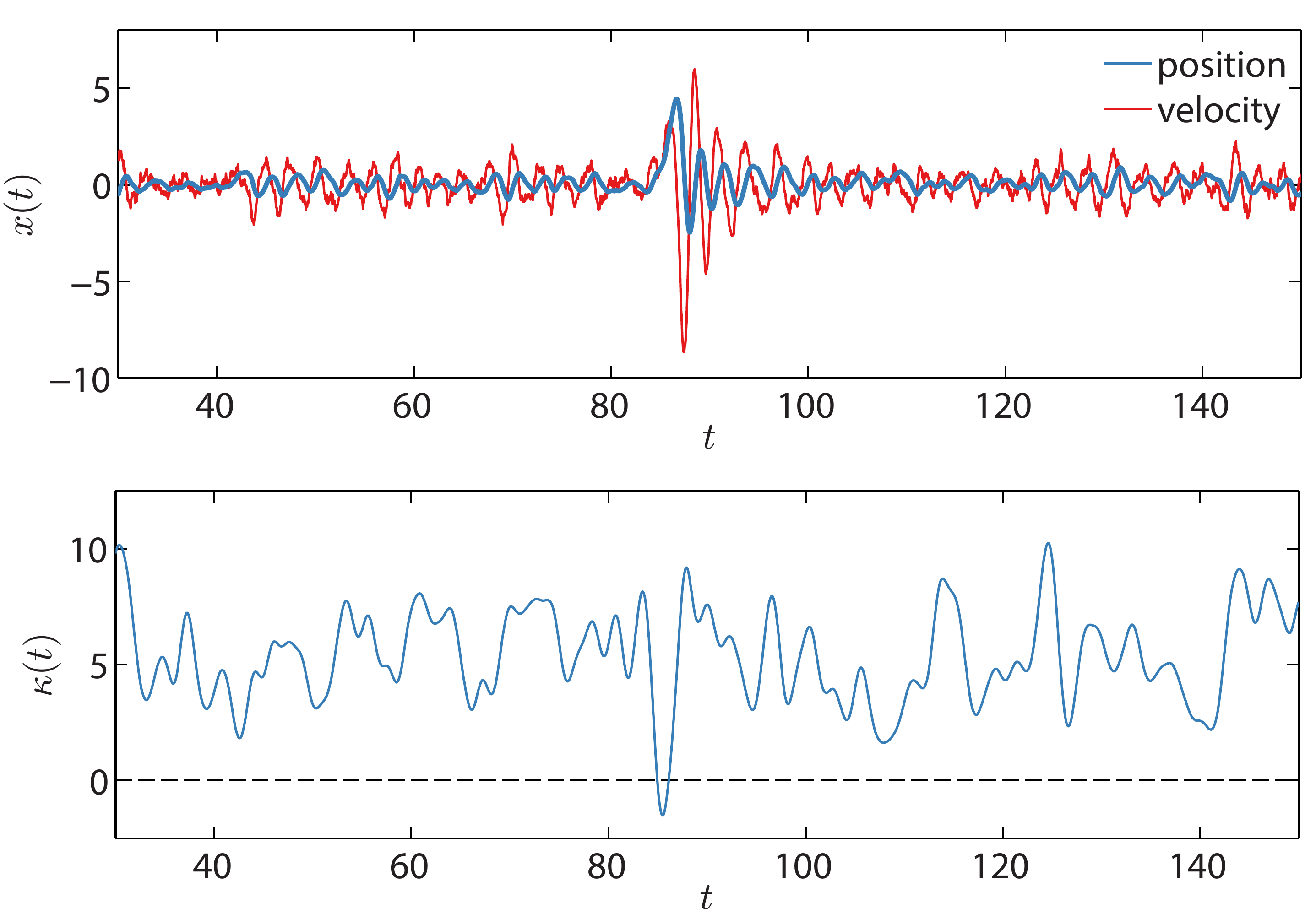}
\caption{Sample realization, demonstrating a typical system response during an intermittent event for position and velocity, for $\eta=-2.27$ ($k=2.2$, $m = 5.0$),  $c= 0.53$.}
\label{fig:samplereal}
\end{figure}

First, in Fig.~\ref{fig:samplereal} a sample system response is shown alongside the corresponding signal of the parametric excitation. This sample realization is characteristic of the typical system response for the  case under study (rare instabilities of finite duration). During the brief period when $\kappa(t)<0$, at approximately $t=85$, the system experiences exponential growth and when the parametric excitation process $\kappa(t)$ returns above the zero level, friction damps the system response from growing out of bounds, returning the system to the stable state after a short decay phase.

\begin{figure}[htb]
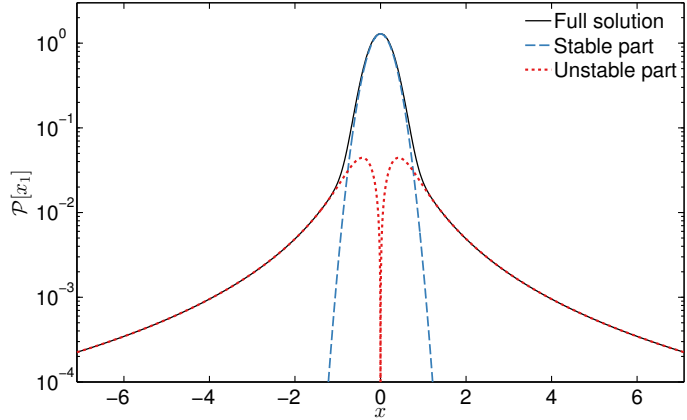

\centering
\includegraphics[width=0.60\textwidth]{{{superPositionPos_k=2.60}}}
\caption{Typical superposition of the stable (in blue) and unstable  (in red) part of the Bayes' decomposition of the full analytical pdf}
\label{fig:superPosition}
\end{figure}

In Fig.~\ref{fig:superPosition}, a typical decomposition of the stable and unstable part of the full response pdf is presented. We emphasize that the tails in the system response pdf are completely described by the term in the Bayes' decomposition that accounts for system instabilities. Whereas the inner core of the pdf is mainly due to the first term in the Bayes' decomposition, which corresponds to the solution in the stable regime. Despite the fact that the system spends practically all of its time in a stable state, the stable part in the expression for the response pdf  contributes essentially zero to the tails of the distribution. The tails, in other words, completely describe the statistics of system instabilities, which are highly non-Gaussian and are described by fat-tailed distributions.

\begin{figure}[htb]
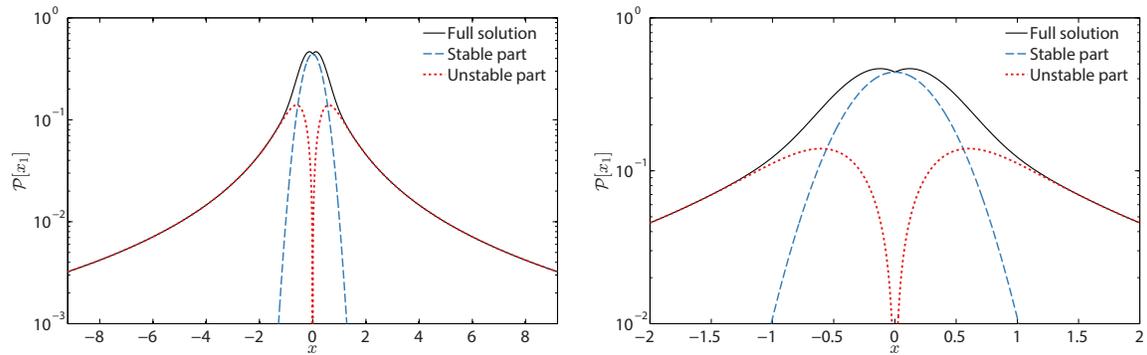

\centering
\includegraphics[width=\textwidth]{{{cuspSuperPosition_k=3.60_position}}}
\caption{Demonstration of the cusp phenomenon, which is prominent in regimes with very frequent instabilities and/or instabilities of long duration, due to approximations made in the unstable regime.}
\label{fig:cusps}
\end{figure}

The analytical results for the system response pdf given in equations~\eqref{eq:pdfpos} and~\eqref{eq:pdfvel} have been derived under a careful set of assumptions. We observe that if the frequency of instabilities is large or if damping is weak (i.e.  extreme events of long duration), or any parameter combination such that unstable part in the Bayes' decomposition contributes significantly to the mean system state, then our analytic expressions for the pdf may display a cusp near $x=0$ (see Fig.~\ref{fig:cusps}). This phenomena is observed, due to the approximation made in Section~\ref{sec:unstable}, where we approximate the conditionally unstable pdf for the position variable in terms of its envelope in equation~\eqref{eq:eq_positio_unstable_oscillator}. Thus, if we are in such a regime, the unstable part of the pdf decomposition contributes a small amount of probability near $x=0$, significant enough to introduce a cusp. This is not the case when the parameters are in a moderate regime, in which case the unstable part of the decomposition will have such a small probability near $x=0$ (essentially zero), so that the Gaussian stable part completely determines the shape of the pdf near the mean state. This phenomenon is more pronounced in the pdf for the velocity $x_{2}$, since the unstable part of the Bayes' decomposition is scaled by a frequency that is smaller than the stable part, which pushes more probability towards the Gaussian core. A minor contributing factor to the severity of this cusp is due to the integral asymptotic expansion made in section~\ref{sec:asymp}; the asymptotic expression in~\eqref{eq:asympunstab} displays sharper curvature near $u_{0}$, when compared to the exact integral in~\eqref{eq:asympunstab}, resulting in a slightly more pronounced cusp.

\subsubsection{Comparison with Monte Carlo simulation}

Here we compare the analytic approximations in equations~\eqref{eq:pdfpos} and \eqref{eq:pdfvel} with Monte Carlo simulations, when the stochastic parametric excitation $\kappa(t)$ is given by an exponential squared correlation function of the form $R(\tau)=\exp(-\tau^{2}/2)$.

\begin{figure}[htb]
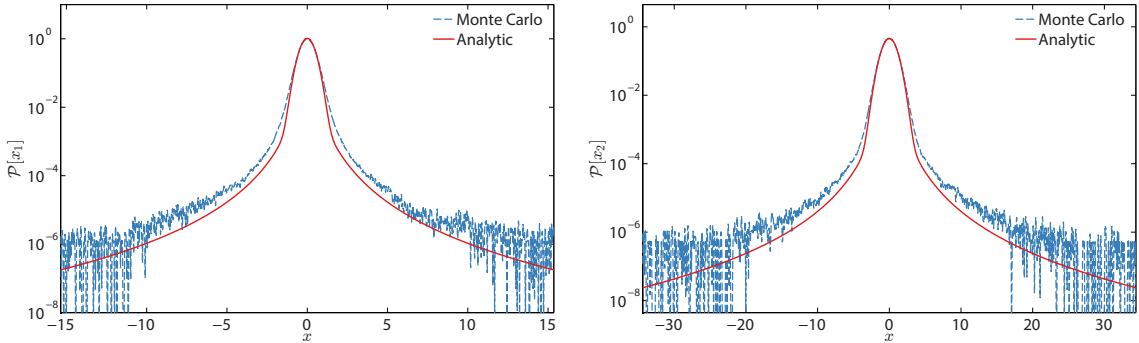

\centering
\includegraphics[width=\textwidth]{{{pdf-w2=5.00_k=1.80_oscno=1.000}}}
\caption{Analytic pdfs~\eqref{eq:pdfpos},~\eqref{eq:pdfvel} compared with results from Monte Carlo simulations for position (left) and velocity  (right), for parameters:  $k=1.8$, $m=5.0$, $c = 0.38$, $\sigma_x=0.75$.}
\label{fig:osck18}
\end{figure}

\begin{figure}[htb]
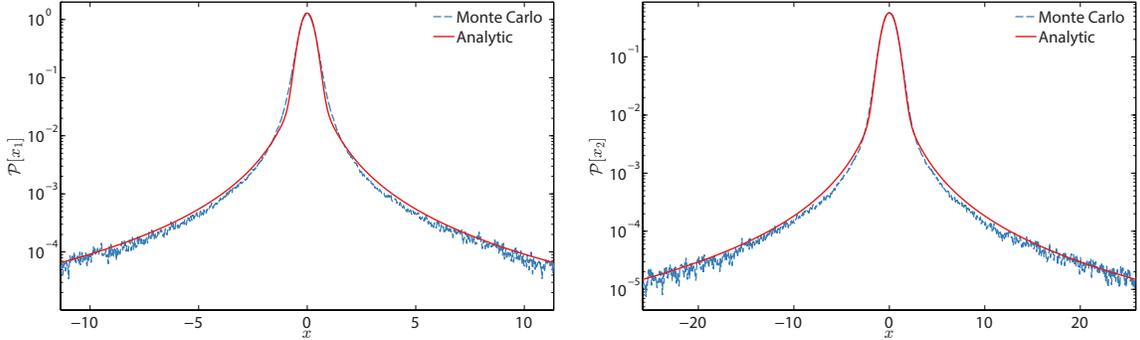

\centering
\includegraphics[width=\textwidth]{{{pdf-w2=5.00_k=2.60_oscno=1.000}}}
\caption{Analytic pdfs~\eqref{eq:pdfpos},~\eqref{eq:pdfvel}  compared with results from Monte Carlo simulations for position   (left) and  velocity  (right), for parameters: $k=2.6$, $m=5.0$, $c = 0.69$, $\sigma_x=0.75$.}
\label{fig:osck26}
\end{figure}

We solve equation~\eqref{eq:harmonic2} using the Euler-Mayaruma method~\cite{higham}, from $t_{0}=0$ to $t_{f}=200$  with step size $\Delta t=2^{-9}$. Realization of the Gaussian process $\kappa(t)$ are generated according to an exact time domain method~\cite{percival92}. To preform simulations we fix the following values for the system parameters $\sigma_{x}=0.75$, $m=5.00$. For the results in this subsection, we furthermore fix the value of friction at $c = 0.38$, which corresponds to an average of one oscillation during the decay phase of an instability. Moreover, for this subsection, only samples after $20$ computational steps are stored and used in calculations, and $5000$ ensembles are used for $k = 1.8$ and $2500$ ensembles otherwise. In Figs.~\ref{fig:osck18},~\ref{fig:osck22}, and~\ref{fig:osck26},  Monte Carlo results for the pdf for position and velocity are compared with the analytic results given in equations~\eqref{eq:pdfpos} and~\eqref{eq:pdfvel}, for various $k$ values, which controls the mean number of instabilities through the parameter $\eta$. Observe that the analytical pdf for position and velocity match the results from Monte Carlo simulations for both extreme events and the main stable response, capturing the overall shape and probability values accurately (the figures for position are plotted out to $50$ standard deviations from the Gaussian core).
\begin{figure}[htb]
\centering
\includegraphics[width=\textwidth]{{{pdf-w2=5.00_k=2.20_oscno=1.000}}}
\caption{Analytic pdfs~\eqref{eq:pdfpos},~\eqref{eq:pdfvel} compared with results from Monte Carlo simulations for position  (left) and velocity (right), for parameters: $k=2.2$, $m=5.0$, $c = 0.53$, $\sigma_x=0.75$.}
\label{fig:osck22}
\end{figure}

\subsubsection{Parameters and region of validity}

Here we quantify the effect of varying system parameters on our analytical approximations. In particular, we consider the effect of both friction $c$ and the frequency of instabilities, controlled by $\eta$, on our results by computing the Kullback-Leibler divergence (KL divergence), which measures the relative information lost by using our approximate analytical pdf instead of the ``true'' pdf from Monte Carlo simulations.

For a more physical interpretation of the effect of the friction parameter, we instead consider the number of oscillations the system undergoes during the decay phase of an instability, i.e. the number of oscillations before the system returns back to the stable regime. To determine the value of damping in term of the number of oscillations $N$ during the decay phase, consider that in Section~\ref{sec:unstable} we derived the approximation $T_{\text{inst}}=(1+{2\bar{\Lambda}}/c)T_{\kappa<0}$, so that the duration of the decay phase is given by $T_{\text{decay}}=({2\bar{\Lambda}}/c)T_{\kappa<0}$. Thus, the average length of the decay phase is $\bar{T}_{\text{decay}}=({2\bar{\Lambda}}/c)\bar{T}_{\kappa<0}$, and the corresponding value of damping that will give on average $N$ oscillations during the decay phase is therefore
\begin{equation}
N=\frac{\omega_{s}}{2\pi}\bar{T}_{\text{decay}}=\frac{\omega_{s}}{2\pi}\frac{2\bar{\Lambda}}{c}\bar{T}_{\kappa<0}\implies c=\frac{\omega_{s}\bar{\Lambda}}{\pi N}\bar{T}_{\kappa<0},
\end{equation}
where $\bar{T}_{\kappa<0}$ is given in equation~\eqref{eq:meantimebelow}
and $\omega_{s}$ in equation~\eqref{eq:fastfreq}.

\begin{figure}[htb]
\centering
\includegraphics[width=\textwidth]{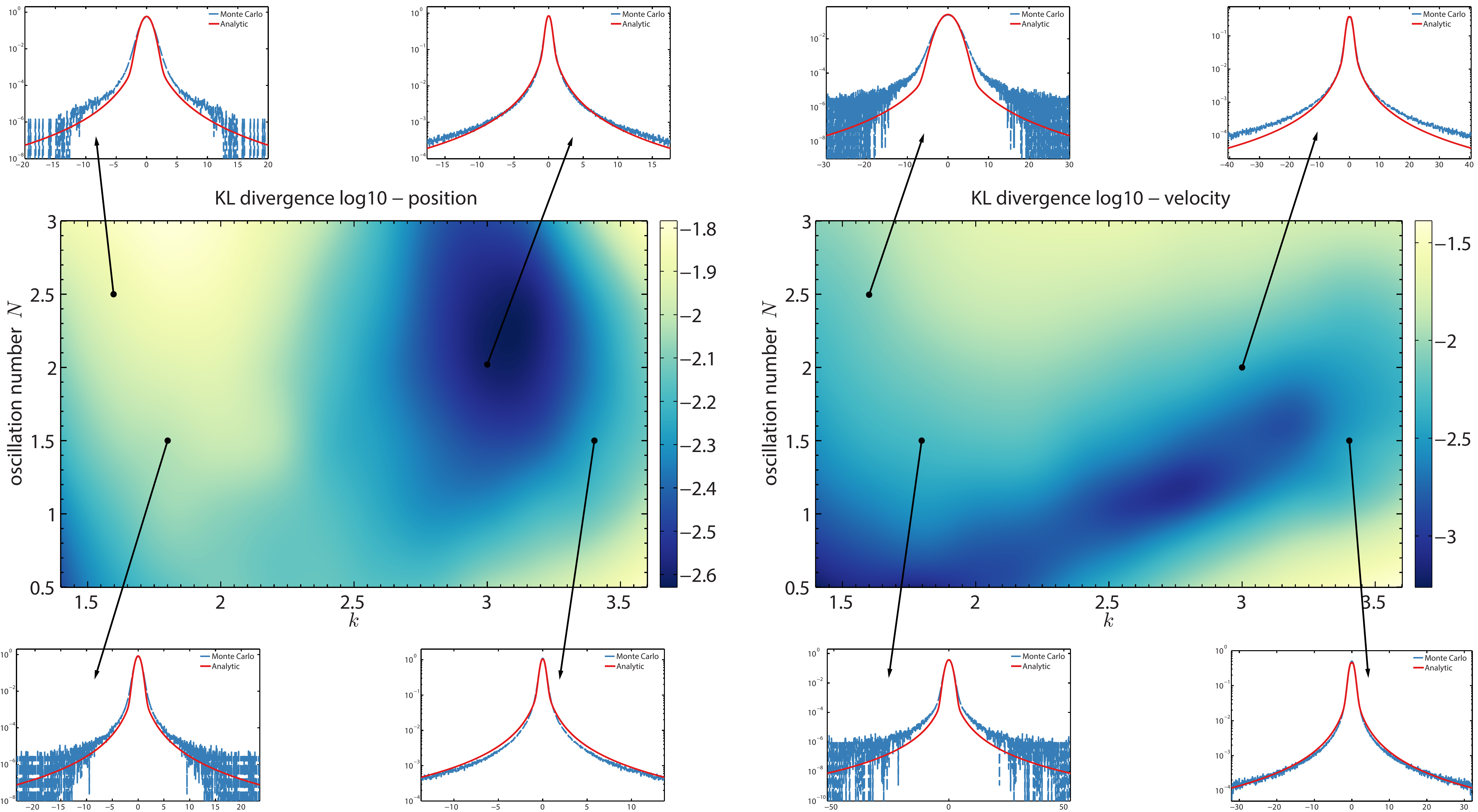}
\caption{KL divergence for position (left) and velocity (right) between analytic and Monte Carlo results for various values of damping (in terms of the number of oscillations $N$ after an instability) and $k$, which controls the number instabilities.}
\label{fig:kldiv}
\end{figure}

In Fig.~\ref{fig:kldiv}, we compute the KL divergence for a range of $k$ and $N$ parameter values. For the simulations we fix $m=5.0$, store values after $40$ computational steps, and use $4000$ ensembles for $k\leq 2.0$ and $2000$ ensembles, otherwise. In addition, we compute the divergence for every half oscillation number from $N=0.5$ to $N=3.0$ and every $k$ value from $k=1.4$ to $k=3.6$, in increments of $0.2$, and interpolate values in between. Overall,  we have good agreement for a wide range of parameters, and even when the analytical results do not capture the exact probability values accurately, we have qualitative agreement with the overall shape of the pdf. The analytical results, deviate when we are in regimes with very frequent instabilities and instabilities of long duration, i.e. small friction values or large $N$ values. This is expected, since in these regimes the instabilities are no longer statistically independent, which we assume in A1. Furthermore, an average of $N=1$ oscillation during the decay phase, shows good agreement with Monte Carlo results, for a large range of frequency of instabilities parameter $\eta$. We note that the KL divergence may be (relatively) large despite the analytic pdf capturing the probabilities in the tails accurately; this is due to the fact that tail events are associated with  extremely small probability values, which the KL divergence does not  emphasize over the stable regime, where the probabilities are large and thus any deviations (e.g. due to a cusp) can  heavily impact divergence values.

\section{Application to  an intermittently unstable complex mode}

Here we present the second application of the method formulated in Section~\ref{sec:methodology}, to that of a complex scalar Langevin equation that models a single mode in a turbulent signal, where multiplicative stochastic damping $\gamma(t)$ and colored additive noise $b(t)$, both specified as Ornstein-Uhlenbeck (OU) processes, replace interactions between various modes. The nonlinear system is given by
\begin{align}
\frac{d u(t)}{d t} &= (-\gamma(t) + i\omega) u(t) + b(t)  + f(t) +\sigma \dot W(t),\\
\frac{d b(t)}{d t} &= (-\gamma_b  + i\omega_b) (b(t) - \hat b)  + \sigma_b \dot W_b(t),\label{eq:turbluntIntModeSys}\\
\frac{d \gamma(t)}{d t} &= -d_\gamma (\gamma(t) - \hat \gamma) + \sigma_\gamma  \dot W_\gamma(t),
\end{align}
where $u(t) \in \mathbb{C}$ physically describes a resolved mode in a turbulent signal and $f(t)$ is a prescribed deterministic forcing. The process $\gamma(t)$ models intermittency due to the (hidden) nonlinear interactions between $u(t)$ and other unobserved modes. In other words, intermittency in the variable $u(t)$ is primarily due to the action of $\gamma(t)$, with $u(t)$ switching between stable and unstable regimes when $\gamma(t)$ switches signs between positive and negative values.

The nonlinear system~\eqref{eq:turbluntIntModeSys} was introduced for filtering of multiscale turbulent signals with hidden instabilities~\cite{Gershgorin20101,Gershgorin201032} and has been used extensively for filtering, prediction, and calibration of climatological systems~\cite{chen_majda_giannakis,branicki_Ger_Majda,branic_majda,Majda_filter,majda_branicki_DCDS}. This system features rich dynamics that closely mimics turbulent signals in various regimes of the turbulent spectrum. In particular, there are three physically relevant dynamical regimes of~\eqref{eq:turbluntIntModeSys}; the three regimes are described by~\cite{branicki_Ger_Majda} (reproduced for completeness):
\begin{enumerate}
\item[R1] This is a regime where the dynamics of $u(t)$ are dominated by frequent, short-lasting transient instabilities, which is characteristic of the dynamics in the turbulent energy transfer range.
\item[R2] In this regime the dynamics of $u(t)$ are characterized by large-amplitude intermittent instabilities followed by a relaxation phase. Here the dynamics are characteristic of turbulent modes in the dissipative range.
\item[R3] This regime is characterized by dynamics of $u(t)$ where transient instabilities are very rare, and fluctuations in $u(t)$ rapidly decorrelate. This type of dynamics is characteristic of the laminar modes in the turbulent spectrum.
\end{enumerate}

We apply the method described in Section~\ref{sec:methodology}, to approximate the probability distribution for the dynamics of $u(t)$ in the special case with no additive noise, i.e. $b = 0$, and no external forcing $f = 0$. This is the simplest case that incorporates intermittency, driven by the state of OU process $\gamma(t)$ (the additive nose term $b(t)$ only impacts the strength of intermittent events). The system we consider is given by
\begin{align}\label{eq:ModeSys}
\frac{d u(t)}{d t} &= (-\gamma(t) + i\omega) u(t) +\sigma \dot W(t),\\
\frac{d \gamma(t)}{d t} &= -d_\gamma (\gamma(t) - \hat \gamma) + \sigma_\gamma  \dot W_\gamma(t),
\end{align}
where intermittent events are triggered when $\gamma(t) < 0$. This application is a good test case of our method for the derivation of analytic approximations for the pdf of intermittent systems, since the dynamics of $u(t)$ are such that it oscillates at a fixed frequency $\omega$, and therefore no approximations need to be made as in Section~\ref{sec:applicationoscillator} regarding the eigenvalues of the growth exponent in the unstable state. For this system, we derive the system response pdf and compare the analytical result for all three regimes R1--R3, but we note that the pdf in regime 3 is nearly Gaussian so the full application of our method is excessive.  Since this application follows in similar vein with the previous application for the parametrically excited oscillator, our treatment here is more sparse, but we make careful note of significant differences.

\subsection{Probability distribution in the stable regime}

Here we derive approximattion of the pdf for $u(t)$ given that we are in the stable regime and, moreover, that we have statistical stationarity. In the stable regime, following Section~\ref{sec:methodologystableregim}, we replace $\gamma(t)$ by the conditional average
\begin{equation}
\bar\gamma|_{\gamma>0} = \hat\gamma + k \frac{\phi(\eta)}{1-\Phi(\eta)}.
\end{equation}
Thus the governing equation in the stable regime becomes
\begin{equation}
\frac{d u(t)}{d t} = (-\bar\gamma|_{\gamma>0} + i \omega)u(t)  + \sigma \dot W(t),
\end{equation}
with the exact solution
\begin{equation}
 u(t) = u(0) e^{ (-\bar\gamma|_{\gamma>0} + i \omega)(t-t_0)} + \sigma \int_{t_0}^t e^{ (-\bar\gamma|_{\gamma>0} + i \omega)(t-s)}\, dW(s).
\end{equation}
Since this is a Gaussian system, we can fully describe the pdf for $u(t)$ in this regime by its stationary mean $\overbar{u(t)} = 0$, and stationary variance
\begin{equation}
 \var({u(t)}) = \var(u_0) e^{-2 \bar\gamma|_{\gamma>0}(t-t_0)} +  \frac{\sigma^2}{2 \bar\gamma|_{\gamma>0}} ( 1 - e^{-2 \bar\gamma|_{\gamma>0}(t-t_0)} ) \to \frac{\sigma^2}{2 \bar\gamma|_{\gamma>0}}, \quad\text{as } t \to \infty.
\end{equation}
Therefore, we have the following pdf for the real part of $u(t)$ in the stable regime
\begin{equation}\label{eq:stablemode}
	\prob[\real({u}) = x  \mid \text{stable}] = \sqrt{\frac{2 \bar\gamma|_{ \gamma>0}}{\pi \sigma^2} }\exp\biggl(-\frac{2\bar\gamma|_{\gamma>0}}{\sigma^2} x^2    \biggr).
\end{equation}

\subsection{Probability distribution in the unstable regime}

In the unstable regime, as in the previous application, we describe the probability distribution in terms of the envelope process. The envelope of $u(t)$ is given by
\begin{equation}\label{eq:modeenvel}
 \frac{d\abs{u}^2}{d t} = 2\real\biggl[ \frac{d u}{d t} u^*\biggr] \implies \frac{d\abs{u}^2}{d t} = -2 \gamma(t) \abs{u}^2 + \sigma^2.
\end{equation}
Following assumption A2 we ignore $\sigma^2$, which does not have a large probabilistic impact on the instability strength, and therefore substituting the representation $\abs{u} = e^{\varLambda T}$ into~\eqref{eq:modeenvel} we get $\Lambda \simeq - \gamma$. Now since $\prob[\Lambda] = \prob[-\gamma \mid \gamma < 0 ]$ we have
\begin{equation}\label{eq:modegrowth}
	\prob_{\Lambda}(\lambda) = \frac{1}{\Phi(\eta)}\prob_\gamma(-\lambda )= \frac{1}{k\Phi(\eta) } \phi\biggl(-\frac{\lambda + \hat\gamma}{k}\biggr), \quad \lambda > 0
\end{equation}
The conditionally unstable pdf will be exactly as in~\eqref{eq:density1}, with the minor modification that the distribution of the eigenvalue of the growth exponent is given instead by~\eqref{eq:modegrowth}. Making this modification we have the following pdf for the envelope in the unstable regime
\begin{equation}\label{eq:unstablemodepdf}
\prob[ u \mid \gamma < 0, u_0 ] = \frac{\pi \log(u/u_0)}{2 k \bar T^2 \Phi(\eta) u} \int_0^\infty \frac{1}{y^2}\phi\biggl(-\frac{y + \hat\gamma}{k}\biggr)  \exp{\bigg(-\frac{\pi}{4 \bar T^2 y^2 } \log(u/u_0)^2\bigg)}\,dy, \quad u > u_0.
\end{equation}
In addition, using~\eqref{eq:instab_duration} we have that the average length of an extreme event be given by
\begin{equation}\label{eq:modeduration}
 \frac{T_\text{decay}}{T_{\alpha<0}} =    -\frac{\hat\gamma - k \frac{\phi(\eta)}{\Phi(\eta)}}{ \hat\gamma + k  \frac{\phi(\eta)}{1-\Phi(\eta)}} \equiv \mu \implies  T_\text{inst} =  (1  - \mu)  T_{\alpha < 0} .
\end{equation}

Finally, to construct the full distribution for the envelope of $u(t)$ in the unstable regime, we need to incorporate the distribution of the initial point of the instability, which is described by the envelope of $u(t)$ in the stable regime, in other words we seek $ \mathcal P[u\mid \gamma < 0] = \mathcal P[u\mid \gamma < 0, u_0]\mathcal P[ u_0]$. The pdf for the envelope of a general process  was described in Section~\ref{sec:envelope}, and using~\eqref{eq:refenvray} we have the following result for the current case
\begin{equation}\label{eq:modeenvelope}
\mathcal P[u_0] =  \frac{4 \bar\gamma|_{\gamma>0}}{\sigma^2} u_0  \exp\biggl(- \frac{2 \bar\gamma|_{\gamma>0}}{\sigma^2} u_0^2\biggr).
\end{equation}

As in Section~\ref{sec:unstable}, we note that the oscillatory character during an instability has be taken into account. However, to avoid the additional integral that would result should this be taken into account by an additional random variable transformation, we instead utilize the same approximation:
\begin{equation}\label{eq:unstablemodeapprox}
	\mathcal{P}[\real(u) = x \mid \gamma < 0] = \frac{1}{2}\mathcal{P}[u = \abs{x} \mid \gamma < 0].
\end{equation}
We will show in the following sections that this approximation compares favorably with direct numerical simulations.

\subsection{Summary of analytical results for the complex mode}

Combining equations \eqref{eq:stablemode}, \eqref{eq:unstablemodepdf}, \eqref{eq:modeenvelope}, and \eqref{eq:modeduration} into the Bayes' decomposition \eqref{eq:bayes} and utilizing the approximation  \eqref{eq:unstablemodeapprox} gives the following heavy-tailed, symmetric pdf for the complex mode, for $x \in \mathbb R$,
\begin{multline}\label{eq:analyticmode}
\mathcal{P}[ \real(u) = x ] = \bigl(1 - (1  - \mu) \Phi(\eta)\bigr)  \sqrt{\frac{2 \bar\gamma|_{ \gamma>0}}{\pi \sigma^2} }\exp\biggl(-\frac{2\bar\gamma|_{\gamma>0}}{\sigma^2} x^2 \biggr) + \\
(1  - \mu) \frac{ \sqrt{2 \pi} \bar\gamma|_{\gamma>0} }{2\sigma^2 k \bar T^2  } \int_0^{\abs{x}}\!\!\int_0^\infty \frac{\log(\abs{x}/u_0)}{y^2(\abs{x}/u_0)}\exp\biggl(-\frac{(y + \hat\gamma)^2}{2 k^2} - \frac{\pi}{4 \bar T^2 y^2 } \log(\abs{x}/u_0)^2 - \frac{2 \bar\gamma|_{\gamma>0}}{\sigma^2} u_0^2 \bigg) \,dy\,du_0.
\end{multline}

Note that in this case, we do not use an asymptotic expansion for the integral in the unstable regime, due to the fact that the resulting expression does not agree as precisely as the integral expansion for the parametrically excited oscillator application. However, the resulting expansion does follow the same decay as the full integral, thus should an asymptotic expansion for the integral be sought, the derivation follows the description given in Section~\ref{sec:asymp}

\subsection{Comparisons with direct Monte Carlo simulations}

Here we compare the analytic results~\eqref{eq:analyticmode} with direct Monte Carlo simulations in the three regimes R1--R3 described in the introduction of the current section. Monte Carlo simulations are carried by numerically discretizing the governing system~\eqref{eq:ModeSys}, using the trapezoidal rule with a time step $\Delta t = 10^{-4}$. We store results after $10$ computational steps and use $1000$ ensembles of time length $t = 600$ (discarding the first $t = 200$ time data, to ensure steady state statistics) for calculating the kernel density estimate for the pdf.

\begin{figure}[htb]
\centering
\includegraphics[width=\textwidth]{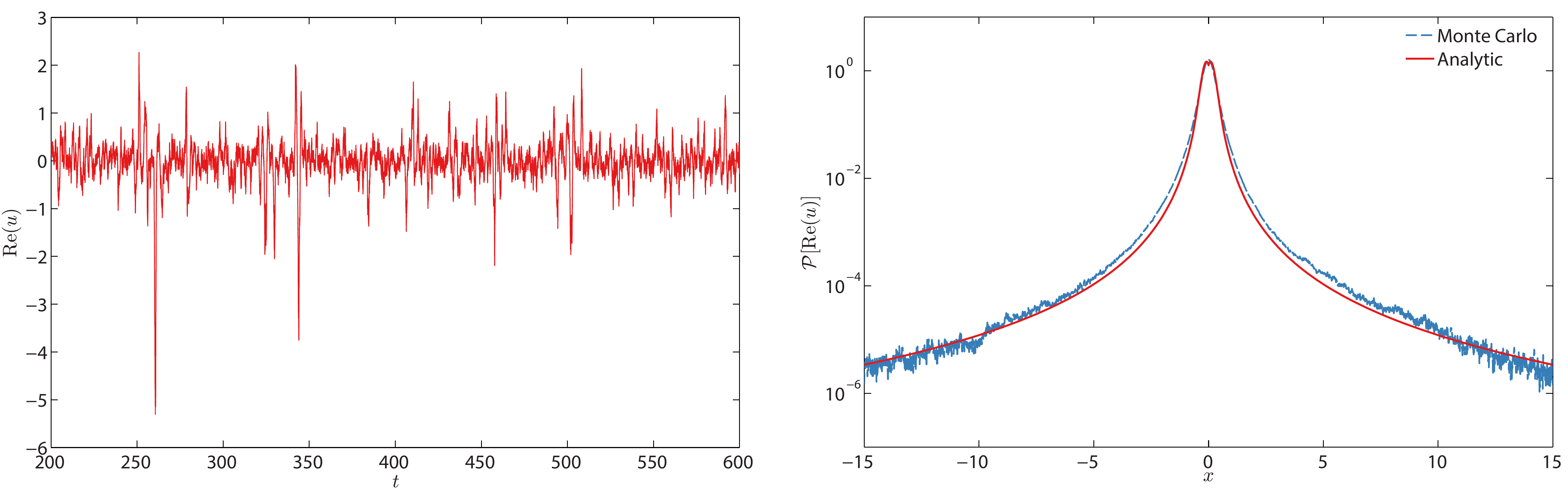}
\caption{Regime 1: Sample path (left) and analytic pdf~\eqref{eq:analyticmode} compared with results from Monte Carlo simulations (right), for parameters: $\omega=1.78$, $\sigma=0.5$, $\hat\gamma = 1.2$, $d_\gamma=10$, $\sigma_\gamma = 10$.}
\label{fig:modereg1}
\end{figure}

\begin{figure}[htb]
\centering
\includegraphics[width=\textwidth]{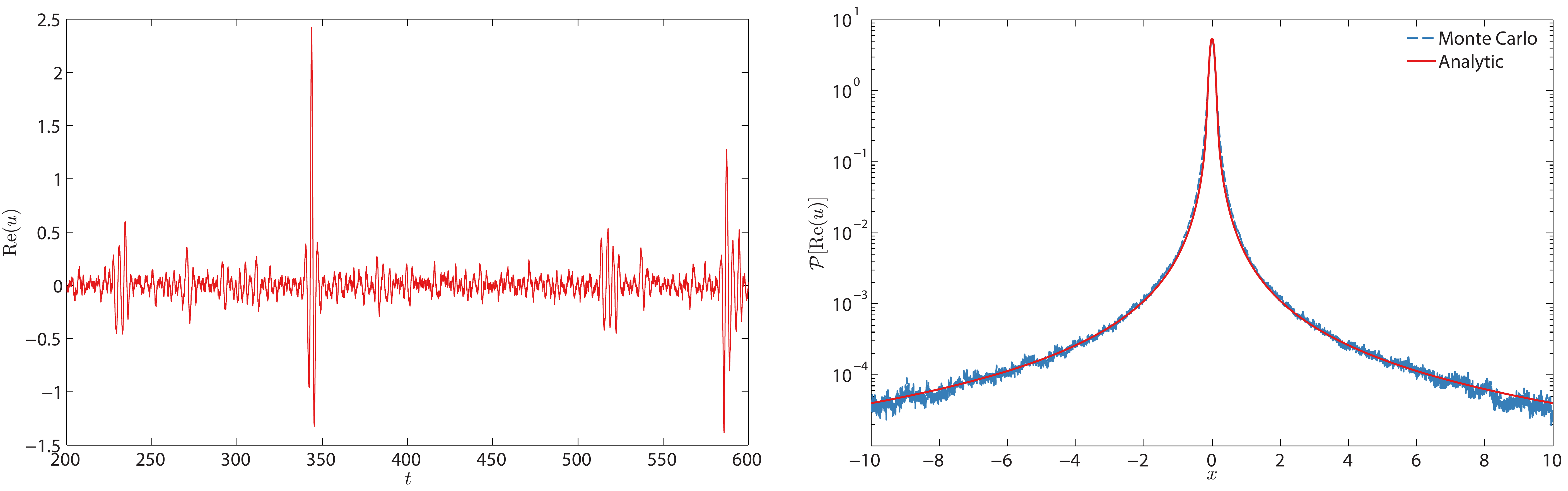}
\caption{Regime 2: Sample path (left) and analytic pdf~\eqref{eq:analyticmode} compared with results from Monte Carlo simulations (right), for parameters: $\omega=1.78$, $\sigma=0.1$, $\hat\gamma = 0.55$, $d_\gamma=0.5$, $\sigma_\gamma = 0.5$.}
\label{fig:modereg2}
\end{figure}

\begin{figure}[htb]
\centering
\includegraphics[width=\textwidth]{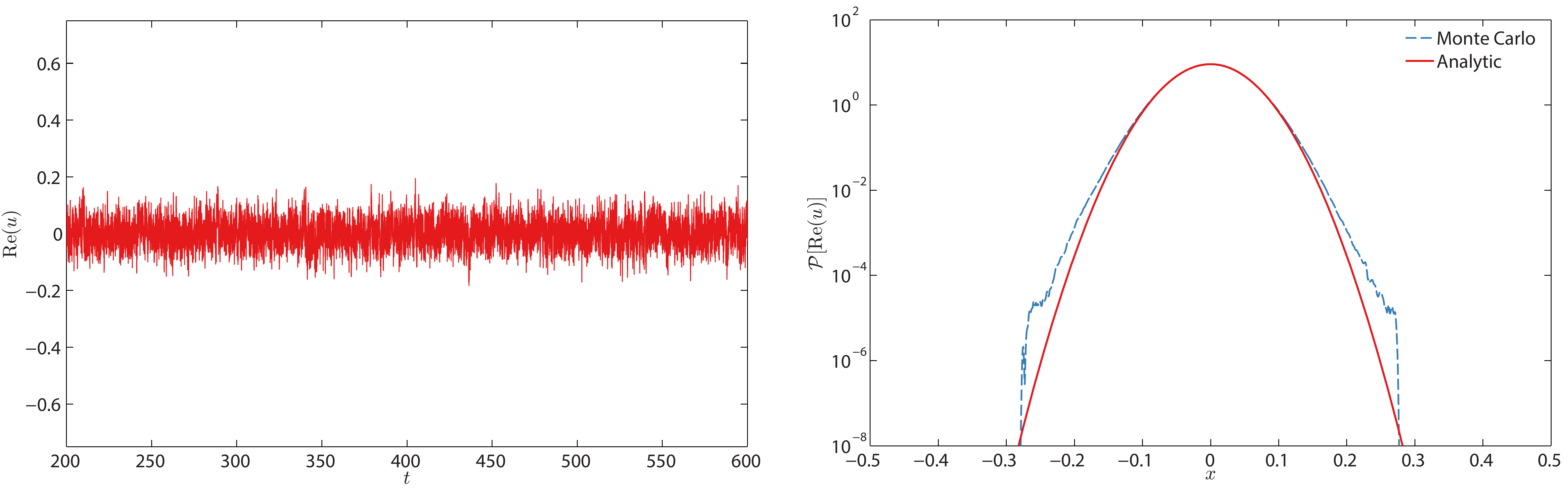}
\caption{Regime 3: Sample path (left) and analytic pdf~\eqref{eq:analyticmode} compared with results from Monte Carlo simulations (right), for parameters: $\omega=1.78$, $\sigma=0.25$, $\hat\gamma = 8.1$, $d_\gamma=0.25$, $\sigma_\gamma = 1$.}
\label{fig:modereg3}
\end{figure}

In Fig.~\ref{fig:modereg1}, the results for Regime 1 are presented alongside a sample realization. As previously mentioned, this regime is characterized 	by frequent short-lasting instabilities. Despite the weak violation of assumption A1, the analytical results are still able to capture the the heavy tails of the response. However, due to the frequency of these short-lasting instabilities and the fact that our analytic results neglects phase information in the conditionally unstable pdf, the cusp phenomena is observed near the mean state. In contrasts, in Regime 2 (see Fig.~\ref{fig:modereg2}), we have large-amplitude instabilities that occur less frequently. Therefore, the analytical results in this regime are able to capture the response extremely accurately even near the mean state, despite no phase information. Again, this is due to the fact that this regime is characterized by less frequent large-amplitude instabilities, which push the conditionally unstable pdf further towards larger magnitude responses, and hence impact the pdf of the conditionally stable regime less severely than in Regime 1. In Fig.~\ref{fig:modereg3} we present the results for Regime 3 for completeness, even though, as previously mentioned, this regime is nearly Gaussian and intermittent events occur very rarely, and under certain parameters almost surely no instabilities will be observed.

\section{Conclusions and future work}

We have formulated a generic method to analytically approximate the pdf of intermittently unstable systems excited by correlated stochastic noise, under a set of assumptions applicable to a broad class of problems commonly encountered in engineering practice. The method developed in this paper relies on conditioning the system response on stable regimes and unstable events according to Bayes' rule, and then reconstruction of the full probabilistic response after the separate analysis of the pdf in the two regimes. Thus, we have demonstrated how the system's response can be decomposed into a statistically stationary part (the stable regime) and an essentially transient part (the unstable regime). We have illustrated how this decomposition accurately captures the heavy-tail statistics that arise due to the presence of intermittent events in the dynamical system's response obtaining a direct link between the character of the intermittent instabilities and the form of the heavy tail statistics of the response.

We have applied the formulated method on two prototype intermittently unstable systems: a  mechanical oscillator excited by correlated multiplicative noise, and a complex mode that represents intermittent modes of a turbulent system, where multiplicative stochastic excitation and additive noise model interactions between various modes.  In both cases, our analytic approximations compare favorably with Monte Carlo simulations under a broad range of parameters. In the mechanical oscillator application, our analysis has furthermore unveiled that intermittency in velocity does not obey the familiar scaling of position with the stable state frequency, and that the conditionally unstable pdf must be scaled by a frequency that accounts for both the stable state frequency (fast) and a (slow) frequency associated with intermittent events. In this case, we have made appropriate approximations to account for this observation, and the resulting pdf compares well with direct simulations. In the complex mode application, we derived the system response pdf, and the analytical approximation compares well with Monte Carlo results for the three regimes commonly observed in turbulent signals.

Future work will include the extension of the method developed in this paper to the study of similar responses for more complex systems, involving more degrees of freedom. In particular, the research will be aimed at applying the developed technique for the probabilistic description of extreme events in ship rolling motion, in nonlinear water waves, and also in the motion of finite-sized particles in fluid flows.

\subsection*{Acknowledgments}

This research has been partially supported by the Naval Engineering Education Center (NEEC) grant 3002883706 and by the Office of Naval Research (ONR) grant ONR N00014-14-1-0520. The authors thank Dr. Craig Merrill (NEEC Technical Point of Contact), Dr. Vadim Belenky, and Prof. Andrew Majda for numerous stimulating discussions.

\section*{Appendix I}

In this Appendix we  provide details regarding the asymptotic expansion of the integral in equation~\eqref{eq:fullden}. According to Laplace's method, this integral will be distributed near the peak value of the term being exponentiated
\begin{equation}\label{eq:asymptheta}
\theta(x,y;x_{0})=-\frac{(y^{2}+m)^{2}}{2k^{2}}-\frac{\pi}{4\bar{T}^{2}y^{2}}\log(x/x_{0})^{2}.
\end{equation}
To find the maxima $y^{*}=\max_{y}\exp(\theta(x,y;x_{0}))$, we differentiate~\eqref{eq:asymptheta} with respect to $y$
\begin{equation}
\frac{\partial\theta}{\partial y}=-2y\biggl(\frac{y^{2}}{k^{2}}+\frac{m}{k^{2}}\biggr)+\frac{\pi}{2\bar{T}^{2}y^{3}}\log(x/x_{0})^{2}.
\end{equation}
To arrive at a closed form analytic approximation for $y^*$, we must determine the relative order of magnitude of $y$. To this end, investigating the terms in the integral in~\eqref{eq:density1}, we note that the maximum at $y_\text{max}$   of $\prob_\Lambda(y)$ will determine the order of magnitude of $y$
\begin{equation}
\frac{d}{dy}\biggl( \alpha y \exp\biggl(-\frac{(y^2+m)^2}{2k^2}\biggr)\biggr) = 0  \implies y_\text{max}^2 \simeq \frac{k^2}{2 m},
\end{equation}
in other words we have $y^2 \sim \mathcal O(k^2/m)$, where $\alpha$ denotes constant terms. Therefore, if $m^2\gg k^{2}$  we may approximate~\eqref{eq:asymptheta} by
\begin{equation}
\frac{\partial\theta}{\partial y}\simeq2y\frac{m}{k^{2}}+\frac{\pi}{2\bar{T}^{2}y^{3}}\log(x/x_{0})^{2}\implies y_{1}^{*}(x;x_{0})=\biggl(\frac{\pi k^{2}}{4m\bar{T}^{2}}\biggr)^{1/4}\log(x/x_{0})^{1/2},
\end{equation}
and if $m^2\ll k^{2}$ we have
\begin{equation}
\frac{\partial\theta}{\partial y}\simeq-2\frac{y^3}{k^{2}}+\frac{\pi}{2\bar{T}^{2}y^{3}}\log(x/x_{0})^{2}\implies y_{2}^{*}(x;x_{0})=\biggl(\frac{\pi k^{2}}{4\bar{T}^{2}}\biggr)^{1/6}\log(x/x_{0})^{1/3}.
\end{equation}

\clearpage
\printbibliography
\end{document}